\documentclass[iop]{emulateapj}
\usepackage{epsfig}
\usepackage{csquotes}
\usepackage{graphicx}
\usepackage{ulem,color}
\usepackage[dvipsnames]{xcolor/xcolor}
\usepackage{dcolumn}
\usepackage{bm}
\usepackage{longtable}
\usepackage{natbib}
\usepackage[bookmarks=true]{hyperref}
\usepackage{xspace}
\usepackage{amsmath}
\def\lesssim{\mathrel{\hbox{\rlap{\hbox{\lower4pt\hbox{$\sim$}}}\hbox{$<$}}}}
\def\gtrsim{\mathrel{\hbox{\rlap{\hbox{\lower4pt\hbox{$\sim$}}}\hbox{$>$}}}}

\newcommand{\bea}{\begin{eqnarray}}
\newcommand{\eea}{\end{eqnarray}}

\newcommand{\bP}{{\bf P}}
\newcommand{\bF}{{\bf F}}
\newcommand{\bU}{{\bf U}}

\newcommand{\prim}{{{\mathbf{P}}}}

\newcommand{\harm}{{\sc Harm3d}\xspace} 
\newcommand{\bothros}{{\sc Bothros}\xspace}

\newcommand{\rpmax}{r_{p}}
\newcommand{\rin}{r_\mathrm{in}}

\newcommand{\msun}{M_\odot}

\def\lambdabar{%
\relax
\bgroup
\def\@tempa{\hbox{\raise.73\ht0
\hbox to0pt{\kern.25\wd0\vrule width.5\wd0
height.1pt depth.1pt\hss}\box0}}%
\mathchoice{\setbox0\hbox{$\displaystyle\lambda$}\@tempa}%
{\setbox0\hbox{$\textstyle\lambda$}\@tempa}%
{\setbox0\hbox{$\scriptstyle\lambda$}\@tempa}%
{\setbox0\hbox{$\scriptscriptstyle\lambda$}\@tempa}%
\egroup
}

\begin{document}

\title{Relativistic Dynamics and Mass Exchange in Binary Black Hole Mini-Disks}

\author{Dennis B. Bowen    $^1$,
        Manuela Campanelli $^1$,
        Julian H. Krolik      $^2$,
        Vassilios Mewes    $^1$,
        and Scott C. Noble        $^3$} 

\affil{$^1$ Center for Computational Relativity and Gravitation, 
  Rochester Institute of Technology, Rochester, NY 14623\\
  $^2$ Department of Physics and Astronomy, Johns Hopkins
  University, Baltimore, MD 21218\\
  $^3$Department of Physics and Engineering Physics,
  The University of Tulsa, Tulsa, OK 74104}

\email{dbb2737@rit.edu}

\begin{abstract}
We present the first exploration of gas dynamics in a relativistic binary black hole system
in which an accretion disk (a ``mini-disk") orbits each black hole.
We focus on 2D hydrodynamical studies
of comparable-mass, non-spinning systems.  Relativistic effects alter the dynamics of
gas in this environment in several ways.  Because the gravitational potential between
the two black holes becomes shallower than in the Newtonian regime, the mini-disks
stretch toward the L1 point and the amount of gas passing back and forth between
the mini-disks increases sharply with decreasing binary separation.  This \enquote{sloshing}
is quasi-periodically modulated at $2$ and $2.75$ times the binary orbital frequency, 
corresponding to timescales of hours to days for supermassive binary black holes.
In addition, relativistic effects add an $m=1$ component to the tidally-driven
spiral waves in the disks that are purely $m=2$ in Newtonian gravity; this component
becomes dominant when the separation is $\lesssim 100$ gravitational radii.
Both the sloshing and the spiral waves have the potential to create distinctive
radiation features that may uniquely mark supermassive binary black holes in the
relativistic regime.
\end{abstract}

\keywords{Black hole physics - hydrodynamics - accretion, accretion disks  - Galaxies: nuclei}

\section{Introduction}

Stellar-mass black hole (BH) mergers were detected for the very first time a little more than a year ago
~\citep{TheLIGOScientific:2016qqj,Abbott:2016nmj,TheLIGOScientific:2016pea}.
This extraordinary discovery marks the beginning of an entirely new field of astrophysics, one
in which experiments like advanced LIGO can be expected to see events similar to GW150914 and GW151226
multiple times per year \citep{2017MNRAS.464.2831O, 2016Natur.534..512B, TheLIGOScientific:2016htt}.

In contrast, mergers of supermassive binary black holes (SMBBHs), remain elusive for the time being.
SMBBHs are expected to be formed during galaxy mergers (see \cite{Khan16,Kelley17} for recent work), but there has long
been uncertainty about how their orbits may evolve toward the relativistic regime and merger
of the BHs \citep{BBR80}.  However, numerous mechanisms to accomplish
this have been studied in recent years.  Dynamical friction against stars and gas or slingshot events
can effectively remove angular momentum and energy from BBHs, and N-body simulations of stellar
loss cone repopulation in galactic nuclei following galactic mergers \citep{Khan11,Vasiliev15,2017MNRAS.464.2301G}
suggest comparatively rapid evolution of SMBBH orbits by these processes.
Fluid mechanisms can also cause orbital compression (see, e.g., \cite{2012AdAst2012E...3D} for a review).
Once the system reaches separations less than $\sim 10^3$ gravitational radii,
gravitational wave (GW) emission should then take take them to coalescence in less than a Hubble time \citep{MP05}.
However, their much lower frequency GW emission requires detectors quite different
from LIGO.  Pulsar Timing Array observations may probe the early inspiral regime of the most massive SMBBHs 
$\left( 10^{9} \msun + \right)$ \citep{SHANNON15} within the next decade, but space 
missions such as LISA \citep{ELISA1,ELISA,2013arXiv1305.5720C} will be necessary to
detect directly the gravitational radiation from the merger proper, and such missions are still very far in the future.

On the other hand, because most SMBBHs are expected to coalesce in gas rich environments 
at the center of galaxies \citep{Cuadra09,Chapon13,Colpi14}, these systems should be excellent targets for 
electromagnetic as well as GW observations.  The difficulty is that our knowledge
of the specific kind of electromagnetic signals to expect remains quite primitive.  To the extent
that the ultimate source of heat in gas near SMBBHs is gravity, the Equivalence Principle
suggests that the total amount of energy available for photon radiation should be directly
related to the mass of gas present during the merger, with the energy per unit mass likely
greatest in the region nearest the BHs \citep{Krolik10}.  Further progress, though,
requires an understanding of the configuration of this gas, which likely depends on
parameters such as the binary mass ratio and the BH spins, both magnitude and
direction, not to mention the availability of gas from the host galaxy's interstellar medium.

Early work suggested that little gas would actually reach the vicinity of merging black
holes despite mass accreting toward it through a circumbinary disk.  The reasoning was
based on two arguments.  The first was that even when gravitational radiation losses
are too weak to force orbital evolution, the binary exerts torques on nearby gas strong
enough to clear out a large cavity in the region within $\simeq 2a$ of the binary (here $a$
is the binary semi-major axis) when the binary mass-ratio is not too far from unity
\citep{ArtymLubow94,ArtymLubow96,DOrazio16}.  It was therefore argued that these same
torques would prevent any mass from
proceeding closer to the binary than the outer edge of that gap \citep{P91}. 
However, a few years ago multi-dimensional numerical simulations of circumbinary disks with internal
stresses showed that streams of gas are, in fact, readily peeled off the inner edges of such disks
\citep{MM08,Shi12,Noble12,DOrazio13}.
More recently, simulations with carefully defined external accretion rates
have shown that essentially all the mass passing through the circumbinary disk is conveyed to the
binary \citep{Farris14,Shi2015}.  \citet{Shi2015} showed in detail how binary torques
acting on streams in the gap can drive gas back out to the circumbinary disk, where a
portion of the streams' mass loses enough of its angular momentum by shock deflection that
it then falls directly to the binary.
The second argument was that once the orbit was tight enough for GW emission to drain
energy from the orbit faster than stresses within the circumbinary disk could drive
inflow, the binary would ``decouple" from the external accretion flow, accepting no
further gas \citep{MP05}.  This too has been undermined by actual simulations.
\citet{Noble12} and \citet{Farris15} showed that, at least for the time required for the binary to shrink by
a factor of a few, accretion can continue at more or less the same rate because the
very fact that orbital evolution is more rapid than stress-driven inflow means that
mass for accretion does not need to be brought in from very far out in the circumbinary
disk.  Thus, recent work has shown that the prospects for finding significant amounts of gas near 
merging SMBBHs are much more favorable than previous efforts indicated.

It then remains to ask what happens to the gas delivered to the SMBBH.  Initial work
focused on matter accreting directly from a circumbinary disk to the BHs during
the few binary orbits immediately preceding merger.  These investigations were
``proof of principle'' calculations, designed to show that gas dynamics and full solution
of the Einstein Field Equations could be done in tandem \citep{2010ApJ...715.1117B,Pal10,Farris11,Bode12, Farris12,Giacomazzo12,Gold14}.  
However, because of their brief duration, the total mass transferred from the circumbinary
disk to the central cavity was relatively small and no formation of individual
disks around the BHs was observed. In contrast, recent
Newtonian simulations (see, e.g.~\citep{Farris14,MunozLai16})
have clearly demonstrated that individual ``mini-disks'' form around each BH over many binary orbital periods.

Therefore, in our approach, we begin with the supposition that the persistent feeding of
material into the domain of the BHs leads to the formation of individual mini-disks around
each BH. The mini-disks grow in extent until tidal forces exerted by the companion destabilizes larger orbits.
This limiting size is often called the ``tidal truncation radius" $r_t$.  Much work has been
done studying such systems in the Newtonian regime
\citep{Paczynski:1977,Papaloizou:1977a,LinPap:1979,ArtymLubow94,Mayama10,VB11,Nelson16}.
The circular orbit data of \citet{Paczynski:1977}, for example, can be fit reasonably well
by the expression $r_{t} = 0.27q^{\mp 0.3}a$ where $\mp$ correspond to the primary and secondary BH
respectively, and $q \leq 1$ is the binary mass ratio
\citep{Roedig:2014}.  Other Newtonian work has shown that tidal torques from the companion can
excite spiral waves that steepen into shocks \citep{Lynden-Bell-Pringle74,SMIS87,Papaloizou95,Ju16,Rafikov16}, 
supplementing the internal stresses due to correlated MHD turbulence.  Similar conclusions
were recently obtained in the context of a binary system where the 
mini-disks are placed around Schwarzschild BHs, but the binary separation 
is taken to be in the Newtonian regime \citep{RyanMacFadyen16}. 

In this paper, we present the {\sl first} hydrodynamic simulations of mini-disks
during both the quasi-Newtonian and GW-dominated inspiral regime of BBHs ($a \lesssim 100M$).
In our approach, we use an approximate general relativistic spacetime \citep{PROJ0,PROJS0},
which accurately describes the dynamics of BBHs during the inspiral phase. 
The inspiral itself is described through the post-Newtonian (PN) equations of motion \citep{Blanchet:2014av}.
Although our implementation of general relativistic effects is valid for any 
BH mass-ratio and can accommodate spins, the work reported here focuses on equal-mass,
non-spinning BHs.  Hydrodynamics is treated in 2D because our goal is to
study how general relativistic effects alter mini-disk dynamics, including tidal
truncation, interactions between the mini-disks, and spiral shocks.
While not incorporating important effects (continuing accretion from the circumbinary disk,
vertical structure, and magnetic fields), the work presented here complements existing simulations of 
these systems 
and is a stepping stone towards the goal of performing 3D general 
relativistic MHD simulations of the entire inspiral to
merger of BBHs with circumbinary disks.

As we will show in detail in this paper, we find that relativistic effects can
create qualitative changes to the mini-disks when the binary separation shrinks to several
tens of gravitational radii or less (hereafter, we quote all distances in
in gravitational units: $r_g \equiv GM/c^2 = M$ when $G=c=1$).  The spiral
waves found in so many Newtonian studies take on an entirely new symmetry.
The shape of the potential through the L1 region changes, allowing substantially
greater mass to find its way out of the mini-disks and into that region.  This
gas, liberated from the mini-disks, continually passes back and forth from
the domain dominated by one BH to that dominated by the other BH and back again, shocking
against the outer edge of the mini-disks at each passage.  In principle, enough mass can
be injected into this ``sloshing" region to create observable photon signals;
in systems with unequal BH masses, the sloshing could even result in
net mass-transfer from one BH to the other.

The remainder of this paper is organized as follows. In Section~\ref{sec:simulation-details}, we
present our hydrodynamics methods, spacetime treatment, and the parameters of
the simulations we have conducted.
In Section~\ref{sec:results} we detail the analysis performed and discuss
our results for the truncation radius, mass-sloshing and spiral shock dynamics.
In Section~\ref{sec:discussion} we discuss the implications of our findings.
Finally, we summarize our work in Section~\ref{sec:conclusion}.

\section{Simulation Details}
\label{sec:simulation-details}

\subsection{Overview}

Our goal is to study the hydrodynamics of mini-disks when the BBH
separation is in the relativistic regime.   As a first step, we
limit the problem to examining 2D inviscid hydrodynamics in the orbital plane.  To
approach a stationary state as rapidly as possible, we choose initial
conditions that are nearly hydrostatic and in which the mini-disks extend
close to (either near or somewhat outside) the tidal truncation radius estimated
from Newtonian orbital mechanics.   In addition, to disentangle purely gravitational
effects from fluid effects, we will complement these hydrodynamic simulations
with test-particle simulations.  The spacetime itself is described by a
high-order PN scheme in which gravity is due entirely to the
two BHs of the binary. In plausible SMBBH circumstances, the gas
mass is a tiny fraction of the binary mass, so its contribution to gravity
should always be negligible.

Throughout the paper, unless otherwise noted, we use
geometrized units in which $G=c=1$.  When used as tensorial indices,
we reserve Greek letters (e.g., $\alpha, \beta, \gamma, \ldots$) for
spacetime indices and Roman letters (e.g., $i, j, k, \ldots$) as
indices spanning spatial dimensions.

\subsection{Hydrodynamics}
\label{sec:mhd-evolution}

We use \harm \citep{Noble09} to solve the equations of general relativistic
hydrodynamics on a background spacetime in flux-conservative
form. These equations amount to conservation of baryon number density
and conservation of stress-energy (see \cite{Noble09} for more
details).  Taken together, they can be written as
\begin{equation}
\partial_t \bU\left(\prim\right) =                                                                       
-\partial_i \bF^i\left(\prim\right) + \mathbf{S}\left(\prim\right) \ ,
\end{equation}
where $\bP$ are the ``primitive'' variables, $\bU$ the ``conserved''
variables, $\bF^i$ the fluxes, and $\mathbf{S}$ the source terms. In
terms of the primitive variables and metric functions they can be
expressed as
\begin{eqnarray}
\bU\left(\prim\right) & = & \sqrt{-g} \left[ \rho u^t ,\, {T^t}_t +
  \rho u^t ,\, {T^t}_j \right]^T \ , \label{cons-U} \\
\bF^i\left(\prim\right) & = & \sqrt{-g} \left[ \rho u^i ,\, {T^i}_t +
  \rho u^i ,\, {T^i}_j \right]^T \ , \label{cons-flux} \\
\mathbf{S}\left(\prim\right) & = & \sqrt{-g} \left[ 0 ,\,
  {T^\kappa}_\lambda {\Gamma^\lambda}_{t \kappa} - \mathcal{F}_t ,\,
  {T^\kappa}_\lambda {\Gamma^\lambda}_{j \kappa} - \mathcal{F}_j
  \right]^T , \label{cons-source}
\end{eqnarray}    
where $g$ is the determinant of the metric, ${\Gamma^\lambda}_{\alpha \beta}$ 
are the Christoffel symbols, and $u^{\alpha}$ are the components of the fluid's
4-velocity. The stress-energy tensor can be written as
\begin{equation}
  T_{\alpha \beta} = \rho h u_{\alpha} u_{\beta} + p g_{\alpha \beta} \ ,
\end{equation}
where $h = 1 + \epsilon + p/\rho$ is the specific enthalpy, $\epsilon$ is the specific
internal energy, $p$ is the gas pressure, and $\rho$ is the rest-mass density. 

The gas's thermodynamics are governed by an adiabatic equation of state
with index $\Gamma=5/3$ and local cooling.  The cooling is introduced via
an explicit source term in the stress-energy conservation equation:
$\nabla_\lambda {T^{\lambda}}_{\beta} = -{\cal L}_{c}u_\beta.$  The fluid
rest-frame cooling rate per unit volume ${\cal L}_{c}$ is determined via
the prescription of \cite{Noble12} in which the gas is cooled at a rate
\begin{equation}
  \mathcal{L}_{c} = \frac{\rho \epsilon}{t_{\rm cool}} \left( \frac{\Delta S}{S_0} + \left| \frac{\Delta S}{S_0}\right| \right) \ , 
\end{equation}
where $\Delta S \equiv S - S_0$, and $t_{\rm cool}$ is the cooling
timescale. This prescription cools the gas to the initial entropy
$S_0$ in regions of increased entropy on a timescale $t_{\rm cool}$
usually set to the orbital period of the fluid element. The
cooling time is set differently in each of four distinct regions; one
for the circumbinary region, one for each mini-disk, and one for the cavity
between the mini-disk and circumbinary regions. We define the
circumbinary region as the full azimuthal extent of an annulus
extending in radius from $r=1.5a$ out to the end of our numerical
domain, where $r$ is the PN Harmonic (PNH) radial coordinate
\citep{Blanchet:2014av} with origin at the binary center-of-mass and $a$ is the
binary separation. In this region the cooling timescale is the local
Keplerian orbital period about the total mass, $t_{\rm cool} = 2 \pi
\left(r+M\right)^{3/2} / \sqrt{M}$, where $M$ is the total BH
mass. The mini-disk regions are defined by the areas within which
$r_{i} \leq 0.45 a$, where $r_{i}$ is the PNH radial distance from the
$i^\mathrm{th}$ BH. In these regions, $t_{\rm cool}$ is again the
local orbital period, but now it is calculated in terms of the local
Boyer-Lindquist (BL) coordinates with respect to the closest BH (see
Section~\ref{sec:hydro-id} for how these are calculated), i.e.,
$t_{\rm cool} = 2\pi r_{BL}^{3/2}/\sqrt{m_i}$. Finally, in the cavity
region between the mini-disks and circumbinary region where there are no
quasi-stable circular orbits, $t_{\rm cool}$ is set to be the
(constant) period of a local Keplerian orbit at $r=1.5a$.

\subsection{Spacetime}
\label{sec:spacetime}
An important part of our prescription is the use of an approximate, analytic BBH spacetime \citep{PROJ0}.
Because it is analytic, numerically integrating the Einstein Field Equations forward in time is unnecessary; avoiding that computational
load allows us to follow inspiraling SMBBHs over the timescales on which gas accumulates
(hundreds of binary orbits).

\begin{figure}[htb]
  \centering{\includegraphics[width=\columnwidth]{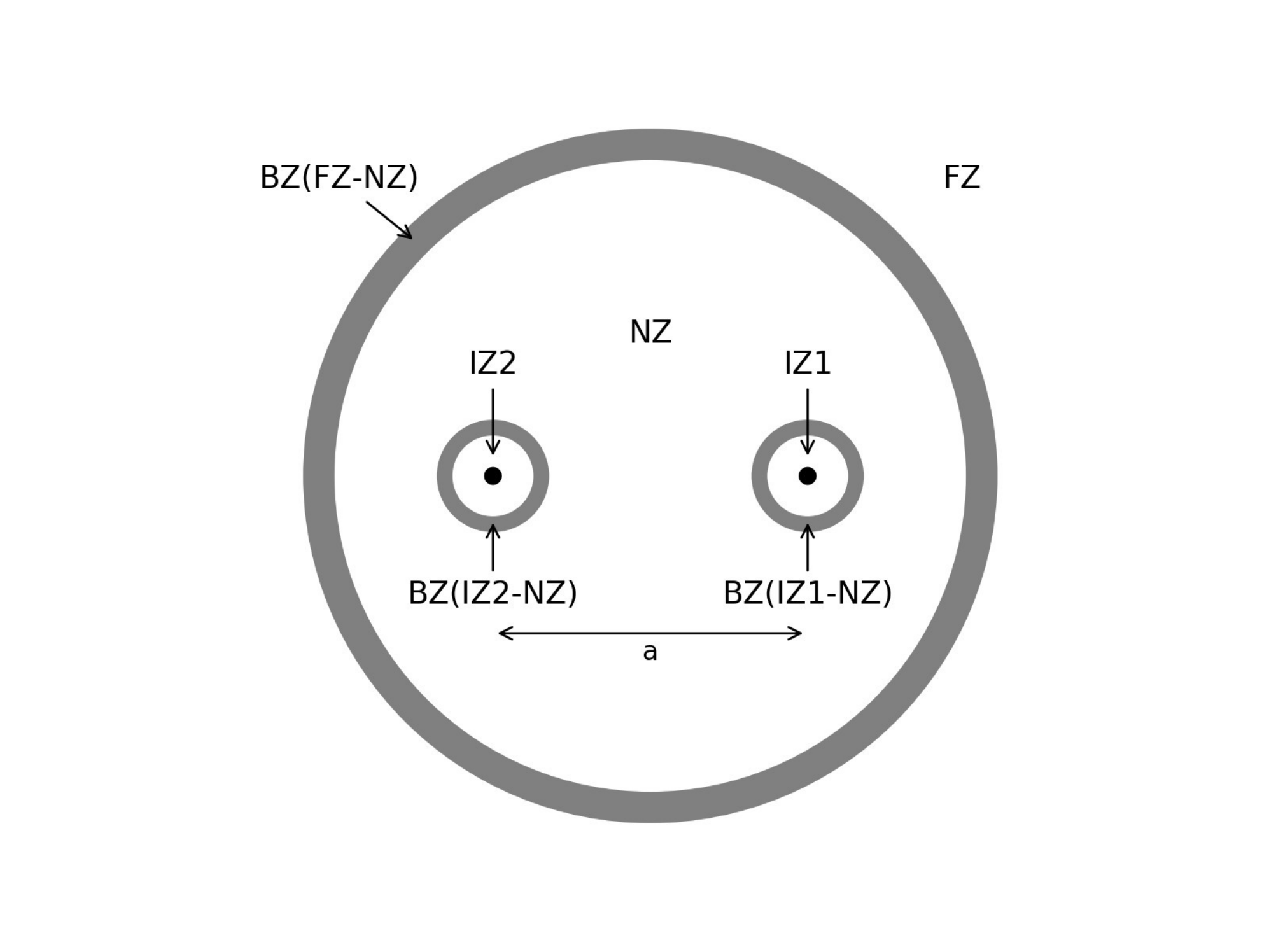}}
  \caption{Schematic representation of our spacetime.  We denote
    the BZs using gray shells and the BHs with black circles. Note
    that these zones are not drawn to scale, and that in practice the
    zones are not perfectly circular as drawn. The full metric is stitched together in
    a weighted sum whose weights are determined using transition
    functions (see \citep{PROJ0,PROJS0} for more details).}
  \label{fig:spacetime-schematic}
\end{figure}

The spacetime is broken into 4 regions:
(i) Inner Zone for BH1 (IZ1), (ii) Inner Zone for BH2 (IZ2), (iii)
Near Zone (NZ), and (iv) Far Zone (FZ). For a schematic representation
of the spacetime see Figure~\ref{fig:spacetime-schematic}.

\begin{figure}[htb]
  \centering{\includegraphics[width=\columnwidth]{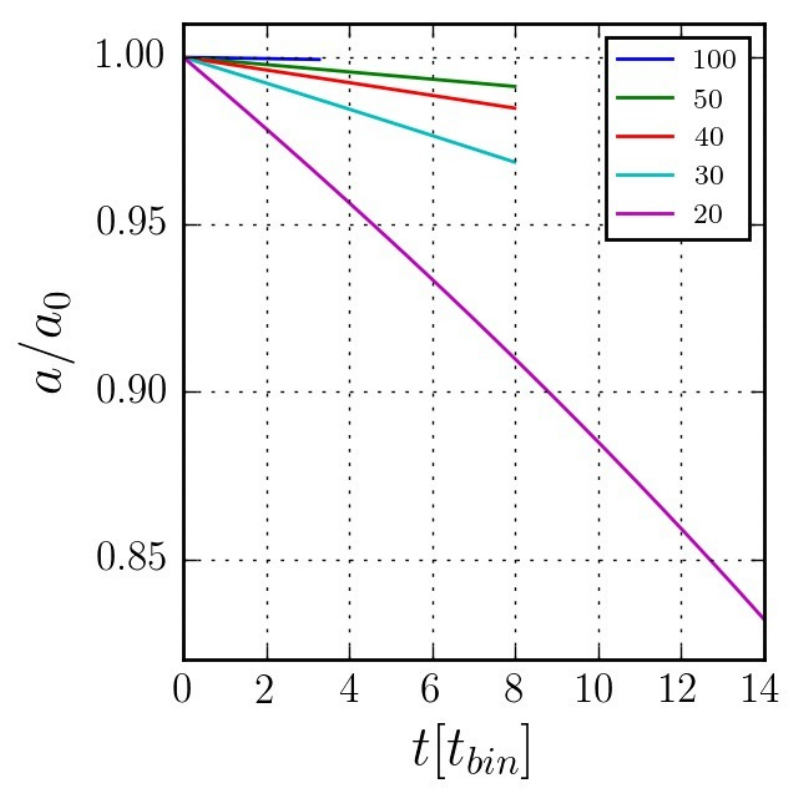}}
  \caption{Binary separation normalized to the initial binary
    separation as a function of binary orbital periods. Note that as
    the separation shrinks the rate of inspiral increases.}
  \label{fig:pn-traj}
\end{figure}

The IZ is defined as the region where the distance to an individual BH
is much less than the binary separation. In these regions the metric
is approximately Schwarzschild, but boosted into the binary center-of-mass
frame and perturbed by the gravity of the other BH. 
The velocity of
the boost is set to be the instantaneous velocity of the BH, as found by the PN 
equations of motion under the quasi-circular approximation \citep{Blanchet:2014av}.  
The equations of motion are accurate to 3.5~PN order and include gravitational 
radiation losses, making the binary's inspiral rate consistent with Einstein's 
equations for the separations explored in this paper.  We note that the spacetime construction is consistent in that 
the same PN equations of motion are used in the NZ spacetime calculation.
The metric in the IZ ensures that our BHs have true horizons by being constructed
via perturbation theory around the Schwarzschild BH solution; it is written in
horizon-penetrating Cook-Scheel harmonic coordinates to avoid coordinate singularities \citep{CS97}.
Farther from the BHs the metric is described by the NZ
metric. This metric encompasses the domain where the separation from
an individual BH is much greater than the BH's mass but still much
less than the gravitational wavelength, $2\pi c/\Omega_{\rm bin}$ in the general
case, $\pi c /\Omega_{\rm bin}$ when $q=1$.
In this domain the metric can be described by PN theory \citep{Blanchet:2014av} in PNH
coordinates.
The mini-disks are contained entirely within
the NZ and IZ domains.  No problems are created at the zone boundaries
because each zone defining the boundary, by construction, shares a region of
common validity with its partners (labeled BZ in the figure).
In these regions the metrics are stitched together through asymptotic
matching. The resultant global spacetime has been shown to satisfy the
Einstein field equations to the expected level of accuracy
\citep{PROJ0}. Finally, the BH trajectories are updated according to
equations of motion accurate to 3.5PN-order; in our largest separation run
($a=100M$), the separation barely changes over the course of the simulation,
while in our smallest separation case ($a=20M$), the orbit shrinks by almost
20\% in only 14 orbital periods (see Figure~\ref{fig:pn-traj}).

\subsection{Initial Conditions}
\label{sec:hydro-id}
The solution for an individual mini-disk in hydrostatic equilibrium in
a binary spacetime is not known. However, solutions for hydrostationary torii orbiting a central mass
are known \citep{1985ApJ...288....1C,dVH03}. These solutions suppose
that the metric has killing symmetries, $\left(\partial_t\right)^{\alpha}$ and
$\left(\partial_{\phi}\right)^{\alpha}$, and that the metric can be expressed
in a spherical coordinate system whose only non-zero diagonal metric component is $g_{t\phi}$.  
Although this is only a crude
approximation to the actual spacetime, and tidal forces do create
significant departures from stationary behavior, it is a feasible
means of constructing an initial state.
Following \cite{Noble12}, we construct the solution for an initially
isentropic accretion disk in hydrostatic equilibrium around an
individual BH as a function of local BL coordinates
$\left(r_{BL},\theta_{BL}\right)$, neglecting the presence of the
binary companion.  We specify the radial location of the inner
edge of the disk ($\rin$) and pressure maximum ($\rpmax$) under the
constraint that $\left(H/r\right) = 0.1$ at the pressure maximum.
For one run ($20M$), $r_{\rm in}$ is inside the innermost stable circular orbit (ISCO); in this case,
we set the initial density and pressure to the floor value for $r < r_{\rm ISCO}$.
Our 2D simulation then evolves the data in the orbital plane, ignoring
the material above and below.  As we show later, the lack of exact hydrostatic
balance leads to a transient, but it decays in 1--3 orbits. 

We relate the local BL coordinates in which the mini-disk initial data
are constructed to the PNH coordinates describing the NZ through
an intermediate Cook-Scheel (CS) harmonic coordinate \citep{CS97}.  For any location
specified in Cartesian PNH coordinates, we calculate the corresponding CS coordinates ($X_{CS}$)
and Jacobian $\frac{\partial X_{PNH}}{\partial X_{CS}}$ as in Appendix~B of
\cite{PROJ0}.  Going the other way, we can express the CS coordinates and
Jacobian $\frac{\partial X_{CS}}{\partial X_{BL}}$ in terms of the
local BL coordinate system \citep{GAL12} through
\begin{eqnarray}
  T &=& t_{BL} + \frac{r_{+}^2 + \chi^2}{r_{+} - r_{-}} \ln \left | \frac{r_{BL} - r_{+}}{r_{BL} - r_{-}} \right | \nonumber\\
  X + i Y &=& \left( r_{BL} - m + i\chi\right)e^{i\phi_{IK}} \sin\theta_{BL}\nonumber\\
  Z &=& \left( r_{BL} - m \right)\cos\theta_{BL}\nonumber\\
  \phi_{IK} &=& \phi_{BL} + \frac{\chi}{r_{+} - r_{-}} \ln \left | \frac{ r_{BL} - r_{+} }{r_{BL} - r_{-}} \right  |
\end{eqnarray}
where $\chi$ is the dimensional spin parameter (zero for this paper),
$r_{\pm} = m \pm \sqrt{m^2 - \chi^2}$, m is the mass of the individual
BH, and the initial BH orbital phase ($\phi_{IK}$) is assumed to be zero.
Combining these two coordinate transformations completes the rule for transforming
quantities between the BL and PNH systems.

The actual simulation is done in a different coordinate system we call
\enquote{warped coordinates} derived from a spherical coordinate system
(see Section~\ref{sec:grid-bound-cond}).  The last stage in initial condition preparation
is therefore to transform the data, originally prepared in BL coordinates, from
Cartesian PNH coordinates (as described above) to spherical PNH coordinates, and
finally to the warped coordinates, i.e.,
\begin{equation}
  X_{BL} \to X_{CS} \to X_{PNHC} \to X_{PNHS} \to X_{WARP}
\end{equation}
where $X_{PNHC}$, $X_{PNHS}$, and $X_{WARP}$ are the Cartesian,
spherical, and warped representation of PNH coordinates,
respectively. Finally, the solution is bi-linearly interpolated onto
the numerical grid.

\begin{figure*}[htb]
  \centerline{\includegraphics[width=\textwidth]{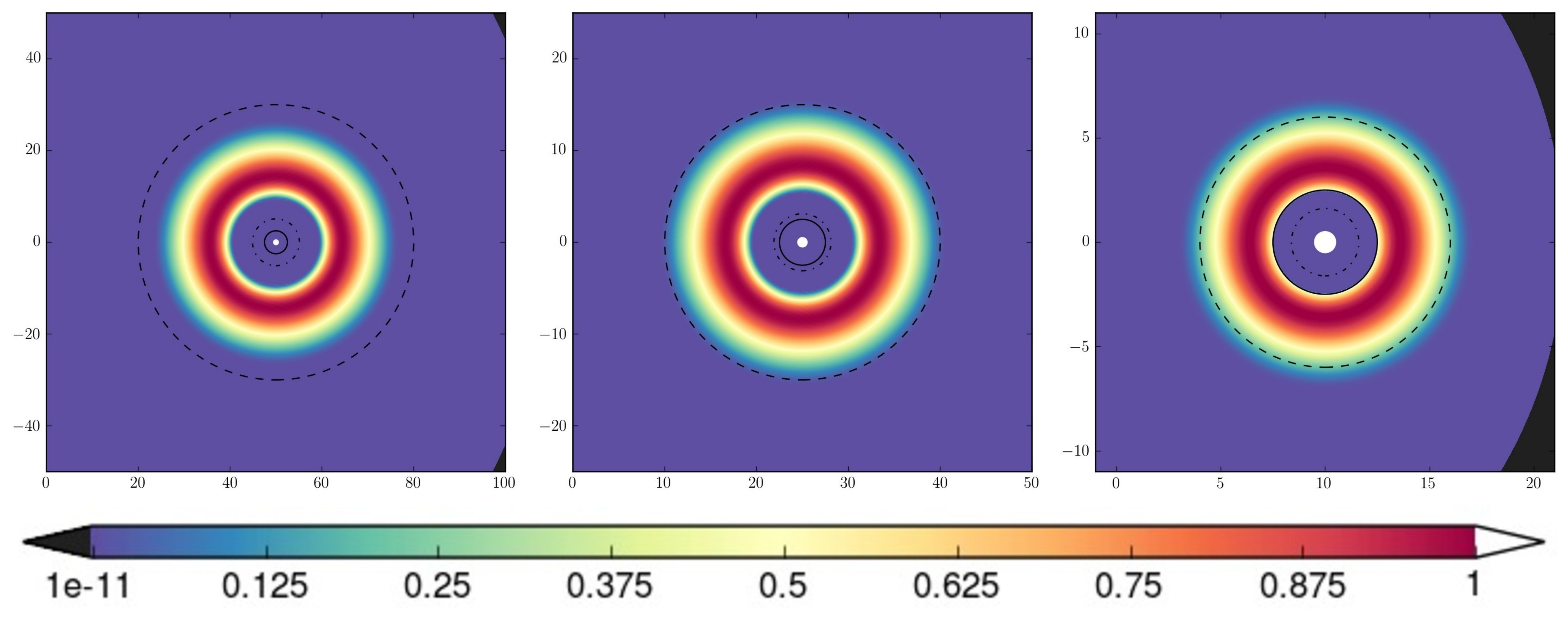}}
  \caption{(Left to right) Linear-scale contours (see color bar) of initial
    density normalized
    to the peak value for the $100M$ Small, $50M$ Small, and $20M$
    runs. For each separation, the BH horizon is represented by a
    white circle, the ISCO by a solid line, the inner edge of the
    IZ-NZ BZ by a dash-dotted line, and the Newtonian truncation
    estimate by a dashed line.}
  \label{fig:hydroID}
\end{figure*}

We performed seven different hydrodynamic evolutions, differing primarily
by initial binary separation.  We tabulate their parameters in
Table~\ref{tab:hydroID}.  For the ``small", $30M$, and $20M$ runs, the mini-disks' initial
outer radii were approximately the Newtonian truncation radius
$(0.3a)$; for the ``large" and $40M$ runs, the initial radii
were larger $(0.4a)$. This distinction permitted us
to explore whether the quasi-steady state obtained
is independent of the initial mini-disk size. The binary separations
range from $100M$ (found to be quasi-Newtonian by
\cite{ZNCZ15}) to $20M$, at which relativistic effects are substantial.
In Figure~\ref{fig:hydroID} we plot the initial
density contours around BH1 (the BH initially on the positive x axis)
for the $20M$ binary separation and ``small'' $50M$
and $100M$ binary separation runs.  For reference, we mark the black
hole horizon, ISCO, the inner edge of the IZ-NZ BZ
(the outer edge of the IZ-NZ BZ lies well outside
the mini-disk), and the Newtonian estimate for the tidal truncation
radius. The majority of the disk mass is located within the IZ-NZ BZ,
well outside the ISCO. For the $50M$ and $100M$ runs, we put in place a
circumbinary disk following the prescription of \cite{Noble12} with $\rin = 3a_0$
and $\rpmax = 5a_0$, with only floor values of density and pressure between the
circumbinary disk and the mini-disks.  We did not observe any substantial inflow from 
the circumbinary into the mini-disks in these runs because, in the absence
of MHD accretion stresses, there is no mechanism to drive accretion
into the central cavity aside from numerical diffusion, and that is kept small
by our spherical grid. Having seen this behavior in the larger separation runs, in the $20M$, $30M$,
and $40M$ runs, we set the initial gas density and pressure everywhere outside the mini-disks
and in the cavity to floor values, eliminating any circumbinary disk. While astrophysical
mini-disks will be influenced by accretion streams from the circumbinary disk,
removing such streams allows us to focus on purely gravitational effects of GR
on the mini-disk structure.

\begin{deluxetable}{ l  c  c  c  c  }
\tablecolumns{5}
\tablecaption{Initial Data Parameters \label{tab:hydroID}}
\tablehead{
\colhead{Run name}     &   \colhead{Initial Separation} & \colhead{$\rin$}      &     \colhead{$\rpmax$}}    
\startdata
$100M$ Large  &   $100M$       &         $10M$        &       $18.5M$   \\
$100M$ Small  &   $100M$       &         $10M$        &       $15M$     \\
$50M$  Large  &   $50M$        &         $6M$         &       $11M$     \\
$50M$  Small  &   $50M$        &         $6M$         &       $9M$      \\
$40M$         &   $40M$        &         $5M$         &       $9M$      \\
$30M$         &   $30M$        &         $3M$         &       $5.5M$    \\
$20M$         &   $20M$        &         $2.5M$       &       $4.2M$    
\enddata
\tablecomments{Initial data parameters used for the hydrodynamic
  runs. Radial coordinates are in the local BL system
  centered on the individual BH. $M$ is the total mass of the binary in all entries.}
\end{deluxetable}

\subsection{Grid and Boundary Conditions}
\label{sec:grid-bound-cond}
Our hydrodynamic simulations are performed in the equatorial plane of a dynamic, double
fish-eye (warped) spherical coordinate system whose origin is at the binary center-of-mass
\citep{WARPED}.  Cells within this coordinate system are spaced
uniformly in numerical spatial coordinates
$\left\{x^{i}\right\}$.  By this means, we are able to focus
resolution in the vicinity of the BHs and rarefy resolution in the
dynamically less interesting portions of the cavity. Near the BHs the
grid is approximately Cartesian, while farther out the grid is spherical.

Given physical PNHS coordinates $\left(r,\phi\right)$, \cite{WARPED}
considered three regions of warping: one for each BH and one for the
region between the BHs. The coordinate transformations from PNHS to WARP take
a different form in each region and are smoothly interpolated (analytically)
to match each other at the regions' boundaries.  The transformations are designed so
that $\Delta r(r)$ has a local minimum at $r=a/2$ and $\Delta
\phi(\phi)$ has a local minimum at the instantaneous azimuthal
positions of the BHs.  Approximately 32 cells span each black
hole horizon in each dimension, a resolution chosen to ensure that the
  near-horizon spacetime is well resolved.
Parameters control various aspects of the transformations.  In particular, it is
important to achieve a smooth transition from nearly-Cartesian cells near the
BH horizons to nearly-spherical cells far from the BHs.  The values we used are stated in
Table~\ref{tab:grids}.  Quantitative expressions for our transformations and the
the parameter definitions may be found in Eqs.~(29-32) of \cite{WARPED}. For convenience,
we also include in Table~\ref{tab:grids} the number of cells within the Newtonian tidal
truncation radius for each binary separation.

\begin{deluxetable*}{ c  c  c  c  c  c }
\tablecolumns{6}
\tablecaption{Warped Grid Parameters \label{tab:grids}}
\tablehead{
\colhead{Initial Separation} & \colhead{$100M$} & \colhead{$50M$} & \colhead{$40M$} & \colhead{$30M$} & \colhead{$20M$} 
}
\startdata
$\delta_{x1}$ & 0.1 & 0.1 & 0.1 & 0.1 & 0.1\\
$\delta_{x2}$  & 0.1 & 0.1 & 0.1 & 0.1 & 0.1\\
$\delta_{x3}$ & 0.25 & 0.18 & 0.21 & 0.21 & 0.18\\
$\delta_{x4}$ & 0.25 &  0.18 & 0.21 & 0.21 & 0.18\\
$\delta_{y3}$ & 0.25 & 0.18 & 0.21 & 0.21 & 0.1\\ 
$\delta_{y4}$ & 0.25 & 0.18 & 0.21 & 0.21 & 0.1\\ 
$\delta_{z}$ & 0.4 & 0.4 & 0.4 & 0.4 & 0.4\\
$a_{x1}$ & 1.7 & 1.67 & 1.64 & 1.60 & 1.50\\
$a_{x2}$ & 1.7 & 1.67 & 1.64 & 1.60 & 1.50\\
$a_{z}$ & 0. & 0. & 0. & 0. & 0.\\ 
$h_{x1}$ & 20. & 20. & 20. & 20. & 20.\\ 
$h_{x2}$  & 20. & 20. & 20. & 20. & 20.\\
$h_{x3}$ & 20. & 20. & 20. & 20. & 20.\\ 
$h_{x4}$ & 20. & 20. & 20. & 20. & 20.\\ 
$h_{y3}$ & 20. & 20. & 20. & 20. & 10.\\ 
$h_{y4}$ & 20. & 20. & 20. & 20. & 10.\\ 
$h_{z}$ & 20. & 20. & 20. & 20. & 20.\\
$s_1$ & 4. & 4. & 0.01 & 0.01 & 0.01\\
$s_2$ & 4. & 4. & 0.01 & 0.01 & 0.01\\
$s_3$ & 4. & 4. & 0.01 & 0.01 & 0.01\\
$b_1$ & 10. & 10. & 10. & 6.5 & 6.5\\ 
$b_2$ & 10. & 10. & 10. & 6.5 & 6.5\\ 
$b_3$ & 40. & 40. & 40. & 40. & 6.5\\ 
$R_{\rm out}$ & $500M$ & $750M$ & $120M$ & $90M$ & $60M$ \\ 
Cell Count & 400 X 400 & 400 X 400 & 300 X 320 & 300 X 320 & 600 X 640\\ 
Cells per Mini-Disk & 25942 & 19258 & 14782 & 14798 & 48532
\enddata

\tablecomments{Parameters of the warped grid used for each initial binary
  separation.  Please see Eqs.~(29-32) of \cite{WARPED} for the
  explicit expressions defining the warped system and the significance
  of these parameters.  For all runs the inner radial cutout is set to
  $1M$.}

\end{deluxetable*}

\begin{figure*}[htb]
  \centerline{\includegraphics[width=\textwidth]{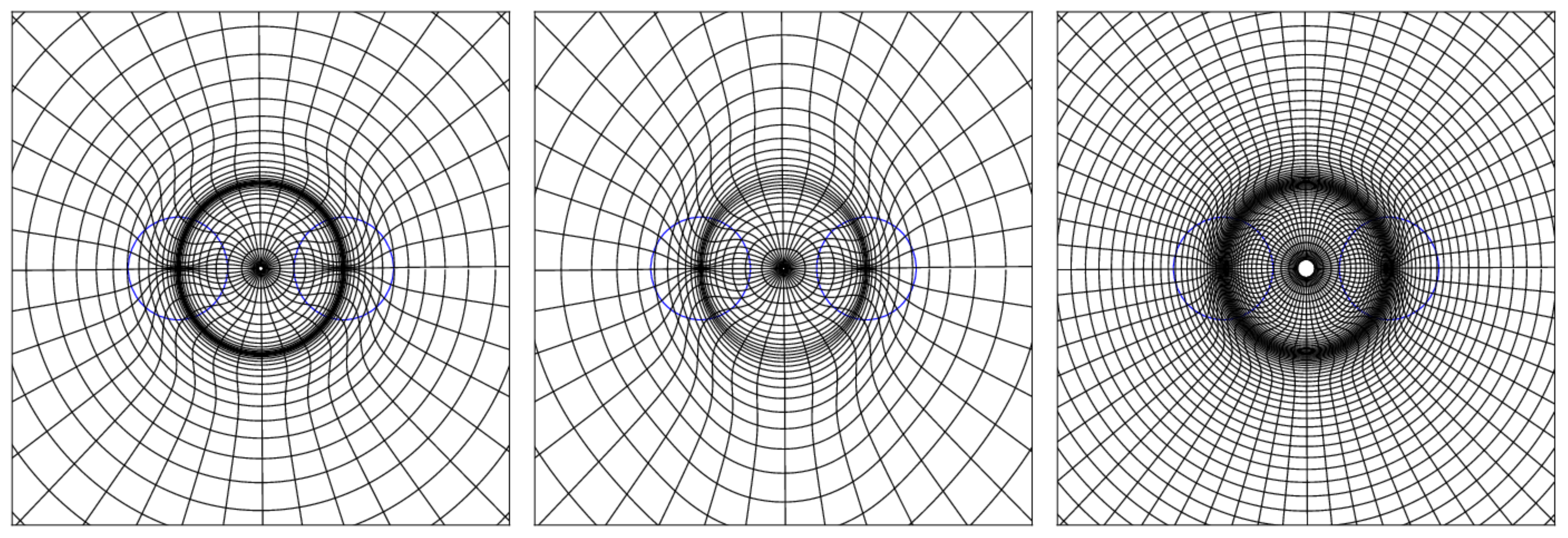}}
  \caption{(Left to right) Grids used for the $100M$, $50M$, and $20M$
    binary separation runs. We plot every tenth grid line within the
    innermost $3a_0 \times 3 a_0$ region of the domain, where $a_0$ is
    the initial binary separation. We show blue circles at $r = 0.3
     a_0$ to illustrate the approximate location of the Newtonian tidal
    truncation radius and the outer edge of mini-disk initial data
   for the ``small'' runs.}
  \label{fig:warpedgrid}
\end{figure*}

The warped coordinates are derived from spherical coordinates, and
therefore possess a coordinate singularity at the origin.
For this reason we excise a sphere of radius $M$ from the
computational domain. In Figure~\ref{fig:warpedgrid} we plot the
grid for the inner $1.5a$ of the $20M$, $50M$, and $100M$ runs.
Throughout the length of our simulations, $\lesssim 1 \% $ of the
initial gas is lost through the origin cutout.

We consistently impose outflow boundary conditions at the radial $x^1$
boundaries, requiring $u^r$ to be oriented out of the domain there.
If at any time the velocity points inward, the value is set to zero,
and we recalculate the primitive velocity.  To enforce our restriction
to 2D, we specify all $x^2$ ghost zones as
copies of the physical cell in the equatorial plane. This effectively
amounts to considering all primitive variables as vertically averaged
quantities that are constant w.r.t. $\theta$. Finally, we apply
periodic boundary conditions in the azimuthal $x^3$ direction and
cover all $2\pi$.

\subsection{Geodesics}

To differentiate between hydrodynamical and purely gravitational
effects we also simulated ensembles of particles orbiting a single BH in
the BBH system. We performed these calculations in \bothros
\citep{Noble07,Noble09, Noble11}, which solves the geodesic equations of motion
\begin{eqnarray}
  \partial_{\lambda} x^{\alpha} &=& N^{\alpha}\\
  \partial_{\lambda} N_{\alpha} &=& {\Gamma^{\kappa}}_{\alpha \eta}N_{\kappa}N^{\eta}.
\end{eqnarray}
Here $N^{\alpha}$ is the particle's 4-velocity, and $\lambda$ is the
affine parameter, or in the case of time-like paths, the proper time
for the particle. We terminate the geodesic if it exits the
central cavity or falls within an ISCO.  \bothros was originally
written for stationary spacetimes, so support for handling dynamic
spacetimes was added for this investigation.  Spatial and temporal
derivatives needed for evaluation of the Christoffel symbols are
computed by centered fourth-order finite differences, using a resolution of
$10^{-6}M$ in space and time for all runs.  The Christoffel symbols
are calculated for every sub-step of the multiple sub-step
Bulirsch-Stoer procedure \citep{1992nrca.book.....P} used to integrate
the geodesics in time.


For a given particle orbiting the non-spinning BH we specify the
initial 4-velocity in local BL coordinates as
\begin{equation}
  u^{\alpha} = \gamma \left( 1, 0, 0, \eta \Omega_K \right)
\end{equation}
where $\Omega_K$ is the Keplerian orbital frequency and $\eta$ is
a parameter selected with uniform probability density within $\left[0.9,1.1\right]$
in order to sample local tidal effects.  This 4-velocity is
then transformed to the $X_{PNHC}$ coordinate system and evolved in
\bothros using the spacetime of Section~\ref{sec:spacetime}. We
perform three runs of 5,376 test particles each at $100M$, $50M$, and
$20M$ binary separations.  The geodesics are launched from $16 \times
16$ uniformly spaced points in $\left(r_{BL}, \phi_{BL}\right)$ space.
We use different ranges of $r_{BL}$ depending on the initial
separation: $r_{BL} \in \left[ 15, 30 \right]M$ for $100M$, 
$r_{BL} \in \left[7.5, 15 \right]M$ for $50M$, and 
$r_{BL} \in \left[3.00, 6.75 \right]M$ for $20M$ binary separations.  
The full $\phi_{BL} \in \left[0, 2 \pi \right]$ extent is used for all runs.  At each
location, 21 geodesics are each launched with a different value of
$\eta$.


\section{Results}
\label{sec:results}

\subsection{Overview}
\label{sec:results_overview}
\begin{figure*}[htb]
  \centerline{\includegraphics[width=\textwidth]{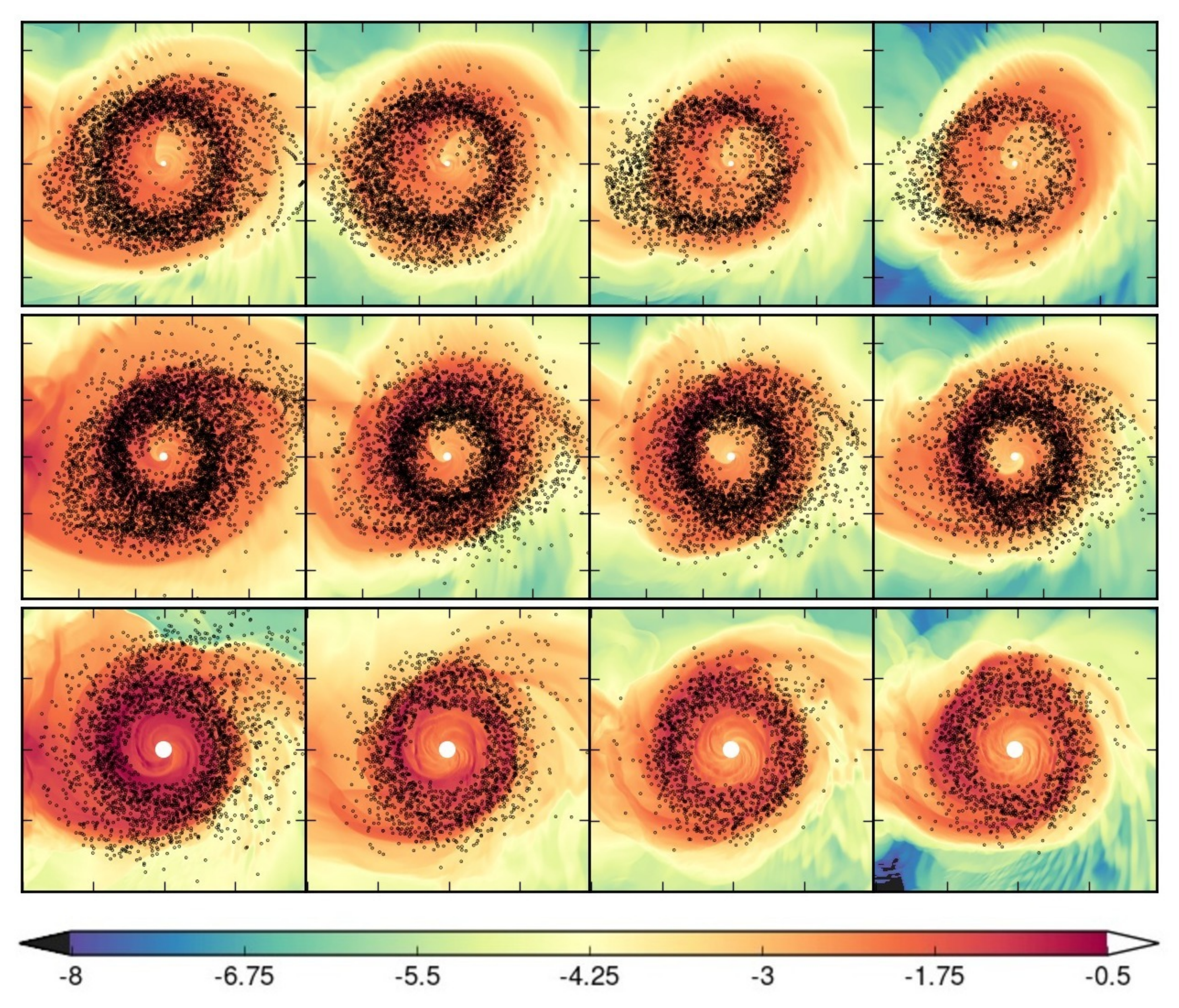}}
  \caption{Logarithmic density contours for the $100M$ Large (top
    row), $50M$ Large (middle row), and $20M$ (bottom row) simulations
    shown underneath particles (black dots) from the test particle
    runs.  Only the region around BH1 to better view the evolution of
    the gas and particles.  BH2 is located off-frame to the left. Columns, from
    left to right, correspond to snapshots taken after 0.5, 1, 1.5,
    and 2 binary orbits.}
\label{fig:rho_snapshots}
\end{figure*}

Our simulations span a range of binary separations from $a \approx 100M$,
where one might expect physics to be quasi-Newtonian, to $a \approx 20M$,
where relativistic effects, including binary inspiral, become very important.
Their general character during the first few binary orbits is illustrated by the
snapshots of the mini-disk around BH1 shown in Figure~\ref{fig:rho_snapshots}.
A casual look suggests there is little variation, either as a function
of time for fixed separation or as a function of separation; a closer
look reveals both significant variability and important trends with
separation.

The first snapshot, at 0.5~orbits, shows the disks at their
greatest extent, as the disks undergo an expansion due to the departures from
hydrostatic balance in our initial conditions caused by the omission of tidal
forces in our approximate hydrostatic balance equations.  However,
one or two more orbits suffice for this transient to decay, letting the
mini-disks achieve their approximate long-term structure.  Once equilibration has
completed, we find that both the hydrodynamic and test particle mini-disks have
settled down to similar stable configurations.

Similarly, although all three separations exhibited show similar initial transients,
they also show a dependence on separation.  In particular, note the greater gas
density on the side toward BH2 (the left side) at smaller separations.
Although not visible in these snapshots, gas is readily
shared back and forth between the two mini-disks in a bar-like region
centered on the L1 point. The fraction of all available gas finding itself in
this region grows sharply with
decreasing separation, rising more than an order of magnitude by $a =
20M$ (see Section~\ref{sec:sloshing} for further details).

Finally, another effect better seen in other figures (see below) is
the development of spiral shocks within the mini-disks. Qualitatively
similar spiral features have been observed in simulations of accretion disks in cataclysmic
variables \citep{Ju16} and BBHs \citep{RyanMacFadyen16}. However, we find that relativistic effects
drive the formation of an $m=1$ mode in addition to the $m=2$ mode
predicted by Newtonian gravity (see Section~\ref{sec:spiralshocks}
for further details).

\subsection{Relativistic Effects in the Potential}
\label{sec:potential}

\begin{figure}[htb]
  \centerline{\includegraphics[width=\columnwidth]{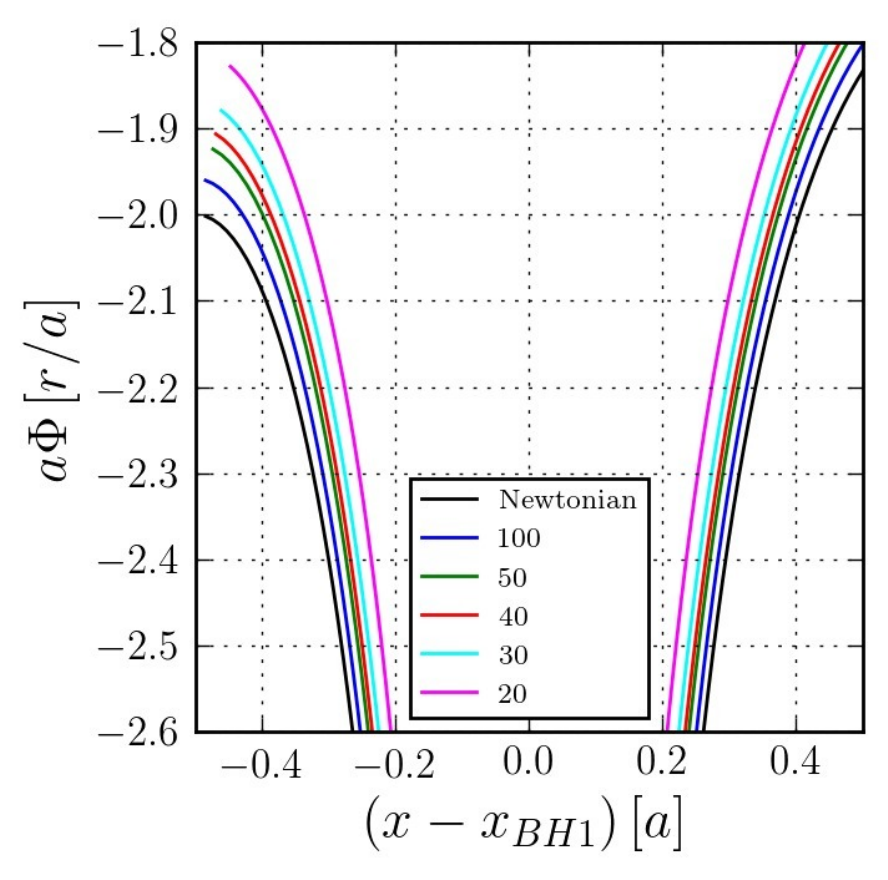}}
  \caption{Binary potential $(\Phi)$ scaled by binary separation
    as a function of distance from BH1 along the line connecting the BHs.
    Distance is measured in units of binary separation and BH2 is located at $-0.5a$.}
  \label{fig:potential_lines}
\end{figure}

\begin{figure}[htb]
  \centerline{\includegraphics[width=\columnwidth]{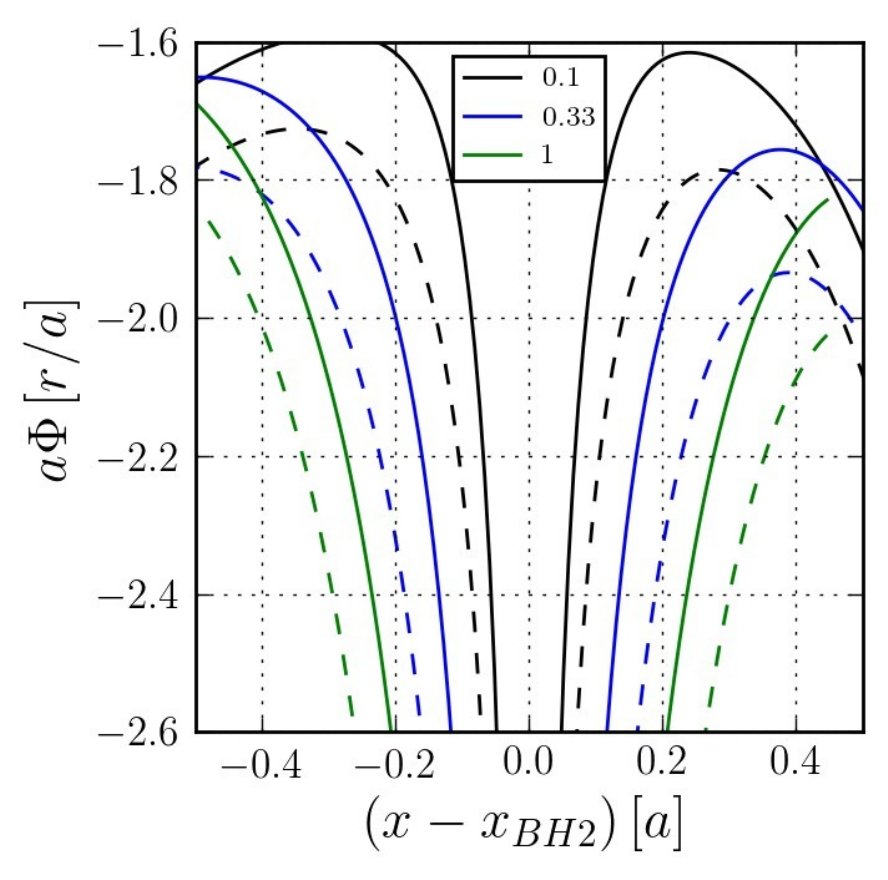}}
  \caption{Binary potential $(\Phi)$ scaled by binary separation
    as a function of distance from the secondary along the line connecting the BHs.
    The primary is located at $+0.5a$.
    All curves are for $a=20M$; distance is measured in units of this separation.
    Solid lines correspond to the spacetime described in Section~\ref{sec:spacetime} and dashed lines
    denote the Newtonian potential. Mass ratio is indicated in the legend.}
  \label{fig:potential_lines_qneq1_secondary}
\end{figure}

\begin{figure}[htb]
  \centerline{\includegraphics[width=\columnwidth]{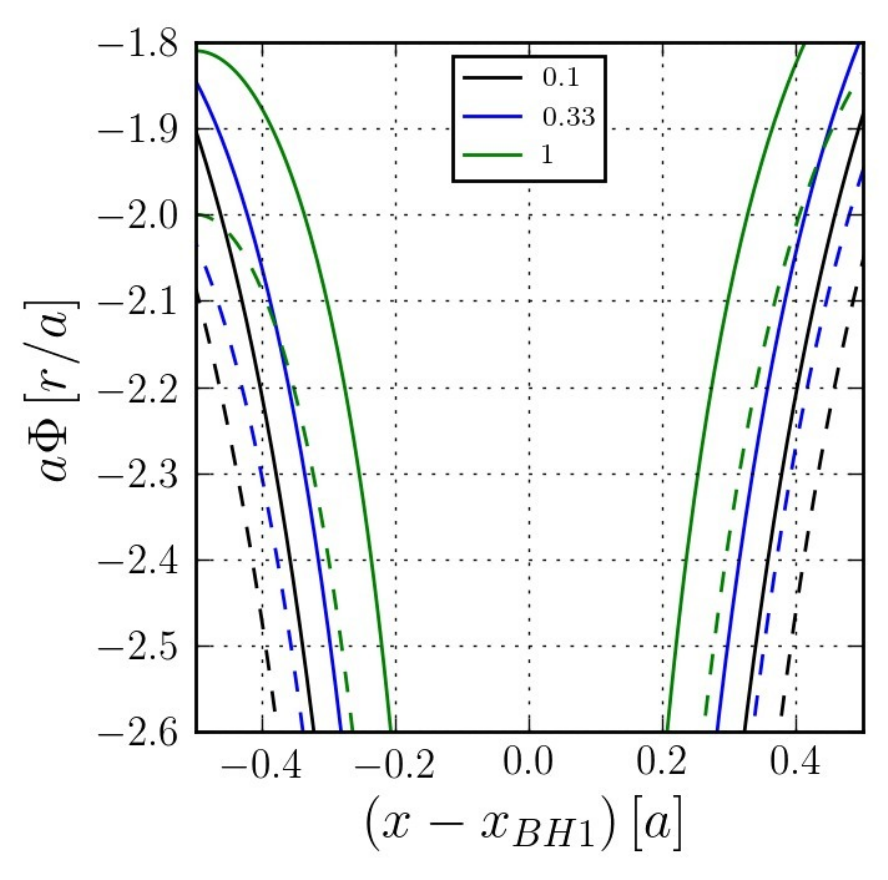}}
  \caption{Binary potential $(\Phi)$ scaled by binary separation
    as a function of distance from the primary along the line connecting the BHs.
    The secondary is located at $-0.5a$.
    All curves are for $a=20M$; distance is measured in units of this separation.
    Solid lines correspond to the spacetime described in Section~\ref{sec:spacetime} and dashed lines
    denote the Newtonian potential. Mass ratio is indicated in the legend.}
  \label{fig:potential_lines_qneq1_primary}
\end{figure}

As discussed in the Introduction, our principal goal is to explore how
mini-disk dynamics may change as relativistic effects become more
important.  The simplest way to highlight how they enter is to study
the lowest-order PN corrections to the metric.  At this
level of approximation, one can isolate the gravitational potential
$\Phi$ in a frame corotating with the binary through the relation
\begin{equation}
  g_{tt} = -\left( 1 + 2\Phi \right).
\end{equation}
This potential combines what in Newtonian language would be called the
genuine gravitational potential with the centrifugal contribution of
the corotating frame.  In Figure~\ref{fig:potential_lines} we plot
$\Phi$, scaled by the binary separation, along the line connecting the BHs
for all separations we simulated.
In Newtonian gravity all the rescaled binary potentials would be identical.
The fact that they differ, and in a fashion that is monotonic with
binary separation, illustrates the relativistic modifications to
Newtonian expectations. The net effect of the PN corrections is to
create shallower potential wells at closer binary separations, particularly
in the secondary's Roche lobe. There, this effect is, in relative terms,
stronger for mass-ratios farther from unity (see Figure~\ref{fig:potential_lines_qneq1_secondary}).
For example, for a separation of $20M$, when $q=1$, the PN potential difference
between the L1 point and the nearest edge of the mini-disk is $0.75 \times$ the Newtonian
potential difference; the corresponding ratios (for the secondary) when
$q=0.33$ and $q=0.1$  are 0.53 and 0.38, respectively. In the primary's
Roche lobe, the trend with mass-ratio is rather weaker, with the ratio
of PN to Newtonian potential difference $\simeq 0.75$--1 and varying
little with $q$ (see Figure~\ref{fig:potential_lines_qneq1_primary}). These
PN changes in the potential will be a recurring theme in our description of the simulations' behavior.

\subsection{Density Distribution and Tidal Truncation of Mini-Disks}
\label{sec:tidaltruncation}

One of our primary goals is to see whether relativistic effects alter the
structure of mini-disks.  To accomplish this, the first step is to demonstrate
that the structures examined are not influenced by transients due to our initial
conditions.  We do so in two ways.  The first is by contrasting simulations in
which the initial conditions differ, and confirming that after the decay of
transients they reach similar states.  The second makes use of time-averaging
over a period later than the $\simeq 2 t_{\rm bin}$ required for transient decay.

\subsubsection{Contrasting Initial Conditions}

To test our results' sensitivity to initial conditions, we ran two versions
of each of the larger separation cases ($100M$ and $50M$).  In the ``small"
runs, the disks were initially filled with matter to a point just inside our
analytic estimate of the tidal truncation radius, $r_t \simeq 0.3a$,
an estimate appropriate to circular-orbit equal-mass Newtonian binaries
\citep{Paczynski:1977,Papaloizou:1977a,ArtymLubow94}.  In the ``large"
runs, in the initial state the disks extended to $\simeq (0.4$--$0.45)a$.

Although it hides departures from axisymmetry, we have chosen the distribution
of mass enclosed within a given radius as a diagnostic of internal structure.
In terms of the conserved rest-mass, this quantity is
\begin{equation}
  M(< r_{BL}) = \int_{r_{\rm min}}^r \, dr^\prime d\phi_{BL} d\theta \rho u^{t}_{BL} \sqrt{-g_{BL}},
\end{equation}
where $r_{\rm min} = 2M_1$.  Because the code data are defined with respect to
warped PNH coordinates, to obtain the data necessary for the integral we first
transformed the code data into the BL system, resolving 4-vector components
with respect to the local BL basis, and then interpolated onto a regular grid
in BL coordinates.

\begin{figure}[htb]
  \centerline{\includegraphics[width=\columnwidth]{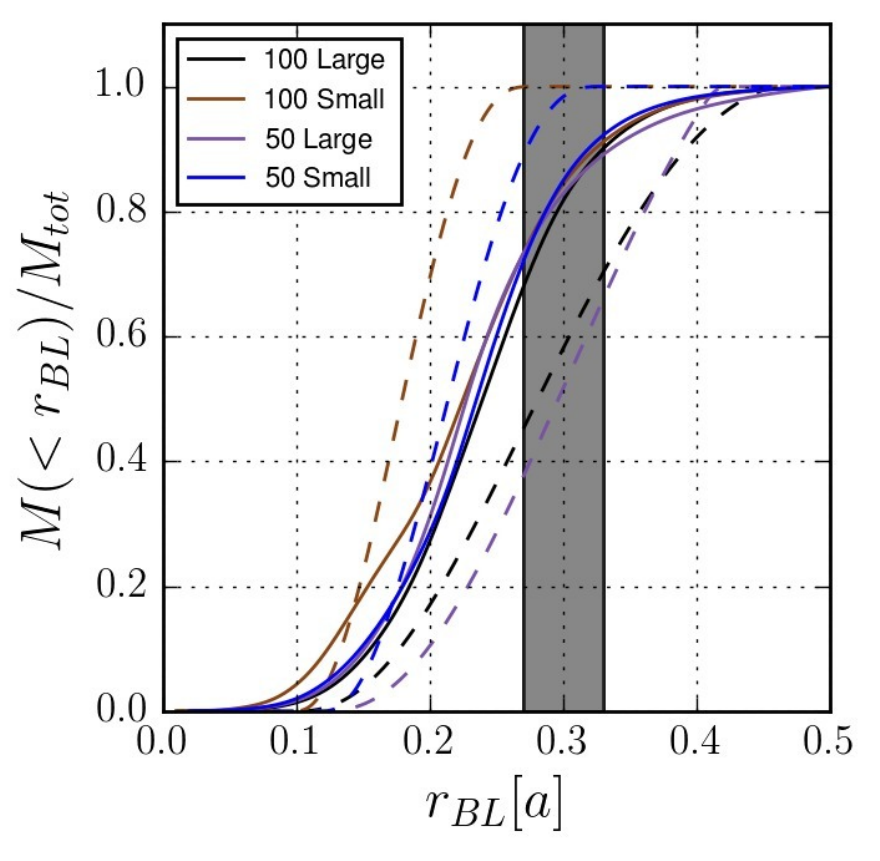}}
  \caption{Mass-enclosed for the $100M$ and $50M$ binary separation
    runs. The solid lines denote the time-averaged data while dashed
    lines correspond to the initial data. The shaded vertical gray
    area denotes the Newtonian prediction range for $r_t$, $(0.27$--$0.33)a$
    \citep{Paczynski:1977,Papaloizou:1977a,ArtymLubow94,Roedig:2014}.}
  \label{fig:mint_a100-50}
\end{figure}

In Figure~\ref{fig:mint_a100-50} we show how well the initial conditions relax
to essentially the same structure by contrasting the large and small initial
mass-enclosed distributions for $100M$ and $50M$ separation with the distribution
averaged over the interval $[2t_{\rm bin},3t_{\rm bin}]$.  Although the $100M$ Small
case differs somewhat from $100M$ Large in the inner quarter of mass even in the later
state of the disk, the remainder of the two $100M$ mass-enclosed distributions are
very close, as are the entire distributions for the Large and Small $50M$ cases.
Because our prescription for the initial condition omits tidal forces, and tidal
forces partially cancel the nearby BH's gravity, the Small cases are
over-pressured; as a result, they expand.  However, tidal forces also prevent
matter from staying near the disk when it lies at distances beyond $r_t$.  For
this reason, the Small cases cease expanding when they reach the tidal truncation
limit, while in Large cases gas beyond $r_t$ does not stay attached to the disk.

On the basis of this success in quickly reaching a state almost entirely independent of initial
conditions, we ran only one case for the $a=20M$, $30M$, and $40M$ separations.  For
$a=40M$, the initial disk extended to $\simeq 0.4a$; for the two smaller separations,
it was filled to only $\simeq 0.3a$.

\subsubsection{Time Averaging}

Before continuing on to other measures of disk structure, it is necessary
to elaborate on our methods of time-averaging.  We chose to begin time-averaging
for all runs at $2t_{\rm bin}$ because the results shown in Figure~\ref{fig:rho_snapshots}
demonstrate that initial transients have almost entirely decayed by this time.
We ended it at $3t_{\rm bin}$ for all cases except $a=20M$ because the
$100M$ simulation stopped at about this time, and we wished to enforce a
consistent procedure on all our cases.  The smallest separation case demanded
special treatment because in it the binary's separation shrinks appreciably
over a single binary orbital period.  In this case, we therefore average successive
intervals of duration $t_{\rm bin}$ beginning at roughly 2, 4, 6, 8, 10, 12, 14, and 15.5 binary
orbits. These correspond to the times at which the binary separation
is $19.5M$, $19.0M$, $18.5M$, $18.0M$, $17.5M$, $17.0M$, $16.5M$, and
$16.0M$. 

\begin{figure*}[htb]
  \centerline{\includegraphics[width=\textwidth]{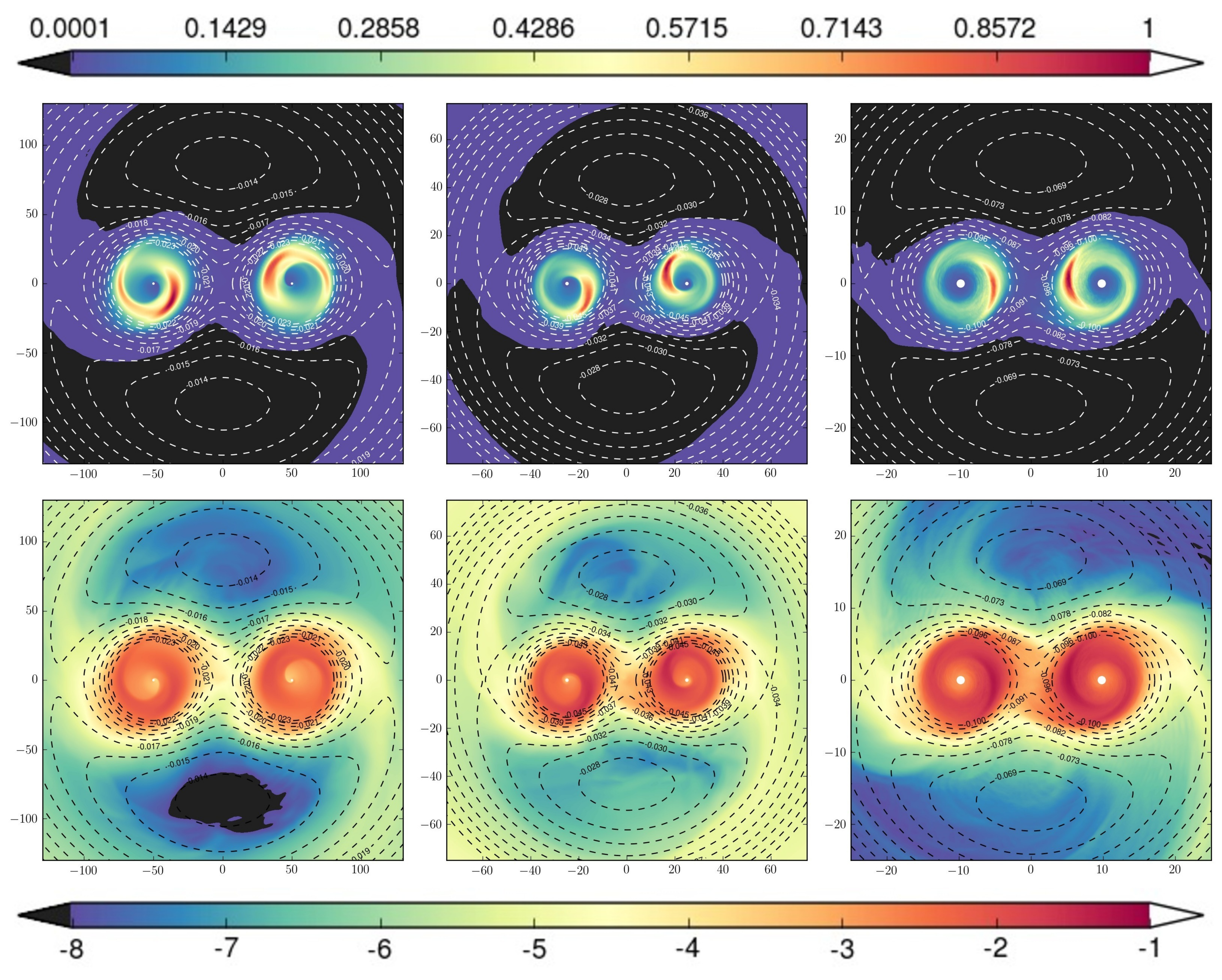}}
  \caption{(Left to right) Time-averaged density contours on a linear
    scale normalized to the peak value (top) and on an unnormalized
    log-scale (bottom) for the $100M$ Large, $50M$ Large, and $20M$
    runs. Time-averaging was done over the third binary orbit for each
    run. Dashed lines overlaid on top of the color density contours
    show the binary's potential, $\Phi$, in the frame corotating with the
    binary.}
  \label{fig:averaged_rho}
\end{figure*}

In Figure~\ref{fig:averaged_rho} we plot the time-averaged density
contours both on a linear and a logarithmic scale, overlaid with
contours of the binary potential evaluated in the frame corotating
with the binary. Several features can be clearly seen in these two
representations. In a linear scale it is apparent that, rather than
being axisymmetric around its BH, the density on the side of each
mini-disk nearer the L1 point is higher than on the opposite side. Interestingly, the
orientation of the density gradient is rotated by $\simeq \pi / 5$
relative to the line between the two BHs when $a=100M$, but this rotation
diminishes with smaller binary separation. On a logarithmic scale, we
see that as the binary separation shrinks,
the relative amount of mass in the zone between the disks rises,
hinting that the density cut-off at the edge of the disk facing the
center-of-mass is becoming less sharp.

\begin{figure}[htb]
  \centerline{\includegraphics[width=\columnwidth]{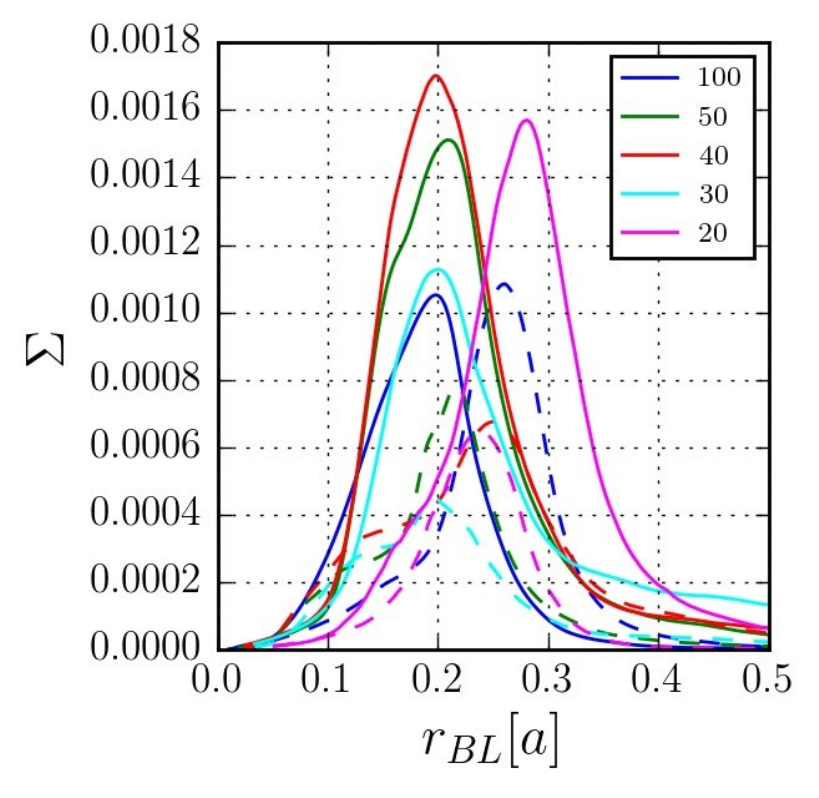}}
  \caption{(Top) Time- and azimuthally-averaged surface density profiles on the near
   (solid) and far (dashed) sides of BH1.  The zero-point of distance is the center
   of BH1 and distance is measured radially outward, whether toward BH2 or away from it.
  }
  \label{fig:gradsigma_lines}
\end{figure}

For a more quantitative view of the time-averaged structure, we study the surface
density,
\begin{equation}
  \Sigma(r_{BL},\phi_{BL}) = \frac{\int \, d\theta \rho \sqrt{-g_{BL}}}{ \sqrt{g_{\phi_{BL} \phi_{BL}}\left(\theta = \pi / 2\right)}} \quad ,
\end{equation}
in local BL coordinates.  The surface density profile of all the BH1 disks is
displayed in Figure~\ref{fig:gradsigma_lines}.  To be more precise, this
figure shows the surface density averaged within a pair of wedges, each
centered on BH1 and stretching $45^\circ$ on either side of the line between
the two BHs.  Several qualitative facts are apparent in this figure.
First, the location of the near-side surface density peak varies little between
$a=100M$ and $a=30M$, but it moves sharply outward when $a=20M$.   On the
other hand, the far side peak position changes little with separation, but
the net sense is to move inward as the separation becomes smaller.  In
addition, although at $a=100M$ the two surface density peaks, near-side
and far-side, are very nearly equal in height, the near-side peak becomes
increasingly dominant as $a$ decreases.

\subsubsection{Quantitative Measures of Tidal Truncation}

\begin{figure*}[htb]
  \centerline{\includegraphics[width=\textwidth]{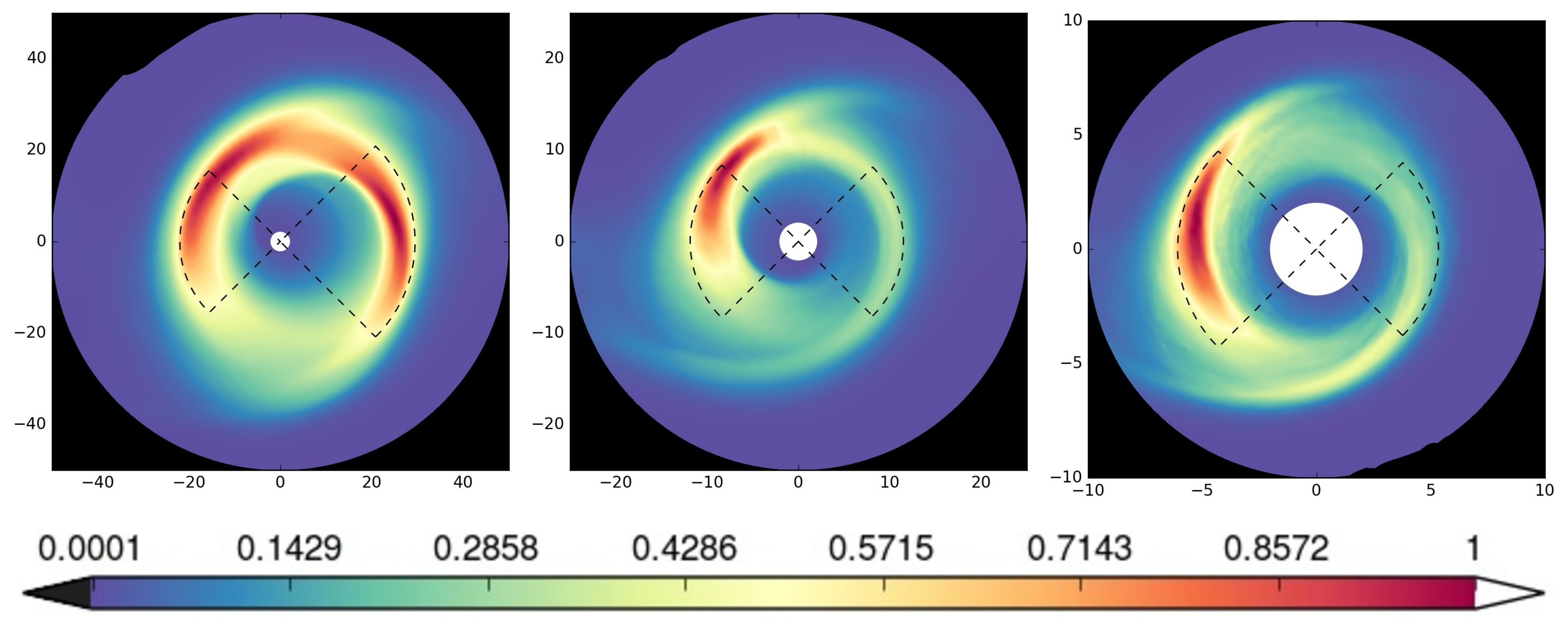}}
  \caption{(Left to right) Time-averaged surface density contours,
    normalized to the peak value, for the
    $100M$ Large, $50M$ Large, and $20M$ runs.  The time-averaging periods used as the same as those used in Figure~\ref{fig:averaged_rho}. 
   The near-side and far-side truncation estimates are represented as dashed wedges.}
  \label{fig:gradsigma_estimates}
\end{figure*}

\begin{figure*}[htb]
  \centerline{\includegraphics[width=\textwidth]{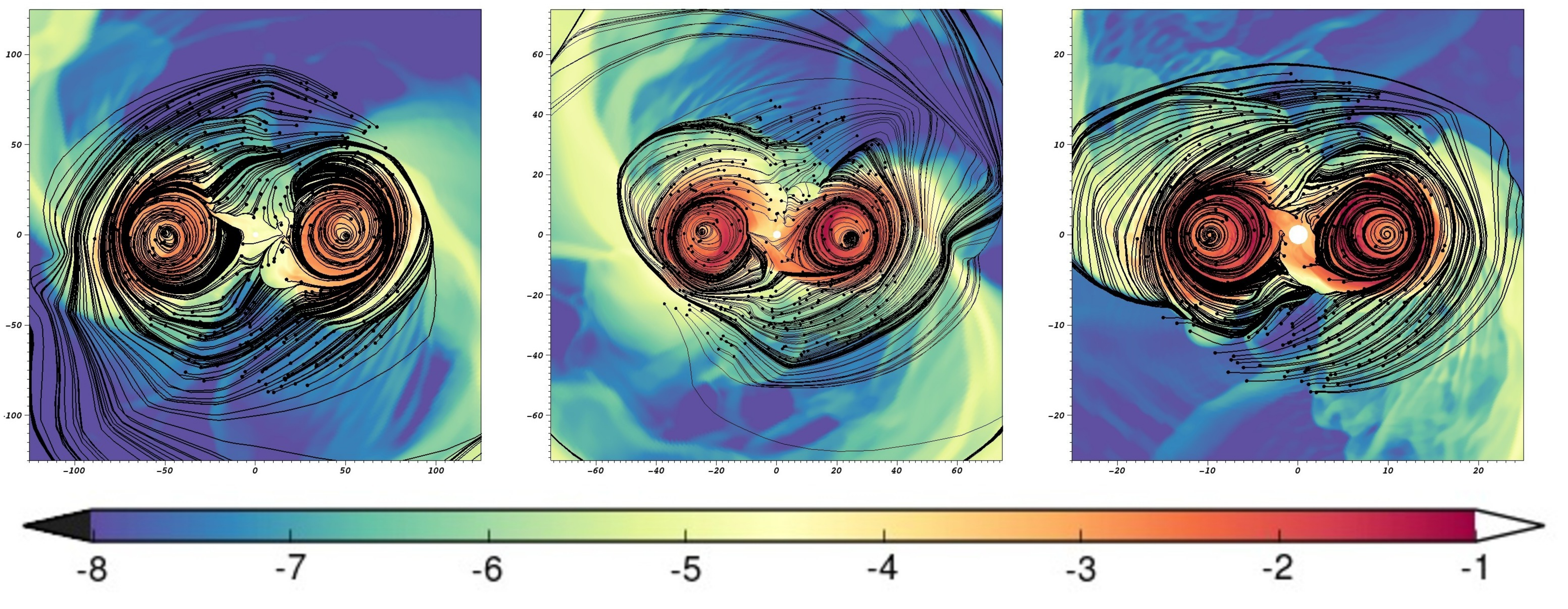}}
  \caption{(Left to right) Logscale density contours for
    the $100M$ Large, $50M$ Large, and $20M$ runs at the start of the
    quasi-steady period with velocity streamlines of the fluid flow in
    black. The start of a streamline is denoted by a black circle.}
  \label{fig:streamlines}
\end{figure*}

The time-averaged surface density profile can be used to determine quantitatively the
edge of the mini-disks in two different ways.  In the first, we define
it as the point where the radial gradient of the time- and azimuthally-averaged
surface density has the greatest magnitude, both on
the side nearest ($0.75\pi \leq \phi_{BL} \leq 1.25\pi$) to and
farthest ($-0.25\pi \leq \phi_{BL} \leq 0.25\pi$) from the binary
companion (see Figure~\ref{fig:gradsigma_estimates}).  In the second,
we search for an outer limit on bound streamlines of the fluid element's velocity
field in snapshots taken during the
quasi-steady state (see Figure~\ref{fig:streamlines}).

In Table~\ref{tab:truncation_radii} we tabulate our estimates for the
edge of the mini-disk around BH1 using both methods on both the side nearest
$(\phi = \pi)$ and farthest from the binary companion $(\phi = 0)$.  Echoing
the behavior of the near-side surface density peak, for separations down to
$a\gtrsim 30M$, the gradient definition yields near-side truncation radii
approximately $0.23a$, while for $a\simeq 20M$, it moves to $\simeq (0.3$--$0.35)a$.
On the far-side, the radii fluctuate between $0.23a-0.30a$ with little trend
as a function of separation.   The streamline truncation radii tell a
different story. They vary more irregularly with binary separation,
ranging between $0.3a-0.4a$ (near-side) and $0.19a-0.24a$ (far-side).  These
contrasting results may be due, in part, to a larger measurement error in determining
the truncation radius from streamlines: contrasting BH1 and BH2 by this measure at
the same time, the radius can differ by as much as $0.05a$.
In a fashion qualitatively similar to Newtonian gravity,
the disks appear to be significantly larger on the near-side than on the far-side,
particularly when defined by streamlines, but also (for $a \lesssim 20M$) when
defined by surface density gradients;
in other words, the truncation ``surface'' is asymmetric, resembling a Roche-lobe shape
more than a sphere.  In addition, particularly as judged by the surface density
gradient, the disks extend farther on the near-side as the binary separation
shrinks.  For larger separations, then, these results are in qualitative agreement
with the Newtonian $r_t \simeq 0.3a$ \citep{Paczynski:1977,Papaloizou:1977a,ArtymLubow94},
but even in that regime it is worth noting the non-circularity of the disks.

\begin{deluxetable}{l c c}
\tablecolumns{3}
\tablewidth{\columnwidth}
\tablecaption{Truncation Measurements \label{tab:truncation_radii}}
\tablehead{
\colhead{$a[M]$}      & \colhead{$grad\left(\Sigma(r,\phi)\right)$} & \colhead{Streamlines} 
}
\startdata
100            & 0.22-0.30          & 0.30-0.24   \\ 
50             & 0.24-0.23          & 0.33-0.17   \\ 
40             & 0.24-0.28          & 0.34-0.24   \\ 
30             & 0.22-0.24          & 0.40-0.19   \\ 
19.5           & 0.31-0.27          & 0.40-0.21   \\ 
19.0           & 0.32-0.26          & 0.33-0.23   \\ 
18.5           & 0.31-0.29          & 0.30-0.22   \\ 
18.0           & 0.31-0.29          & 0.33-0.22   \\ 
17.5           & 0.34-0.28          & 0.39-0.22   \\ 
17.0           & 0.33-0.28          & 0.35-0.19   \\ 
16.5           & 0.35-0.29          & 0.32-0.21   \\ 
16.0           & 0.29-0.27          & 0.34-0.23   
\enddata
\tablecomments{Measured radial extent of the mini-disks on the
  NEAR$(\phi_{BL}=\pi)$-FAR$(\phi_{BL} = 0)$ sides of the disk as
  fractions of the binary separation for the surface density and
  streamline method.}
\end{deluxetable}

\subsection{Sloshing}
\label{sec:sloshing}

\begin{figure}[htb]
  \centerline{\includegraphics[width=\columnwidth]{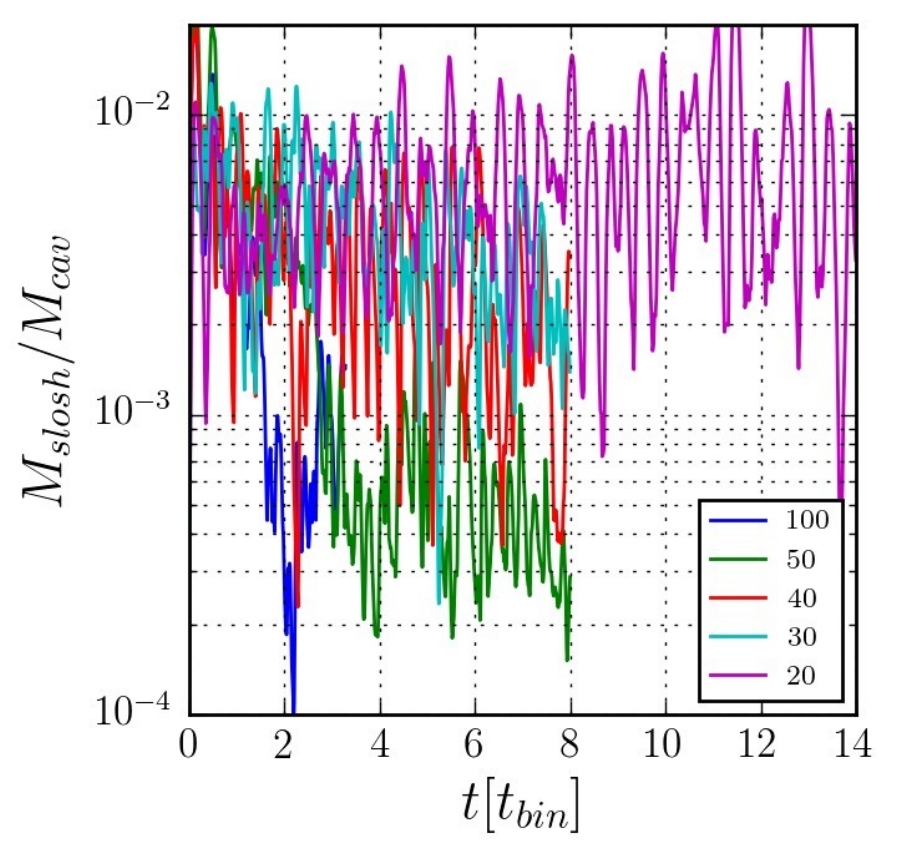}}
  \caption{The mass within the ``sloshing region'' normalized
    to the total mass of the cavity for the $100M$ large, $50M$ large,
    $40M$, $30M$, and $20M$ hydrodynamic runs.}
  \label{fig:sloshing-mass}
\end{figure}

\begin{figure*}[htb]
  \centerline{\includegraphics[width=\textwidth]{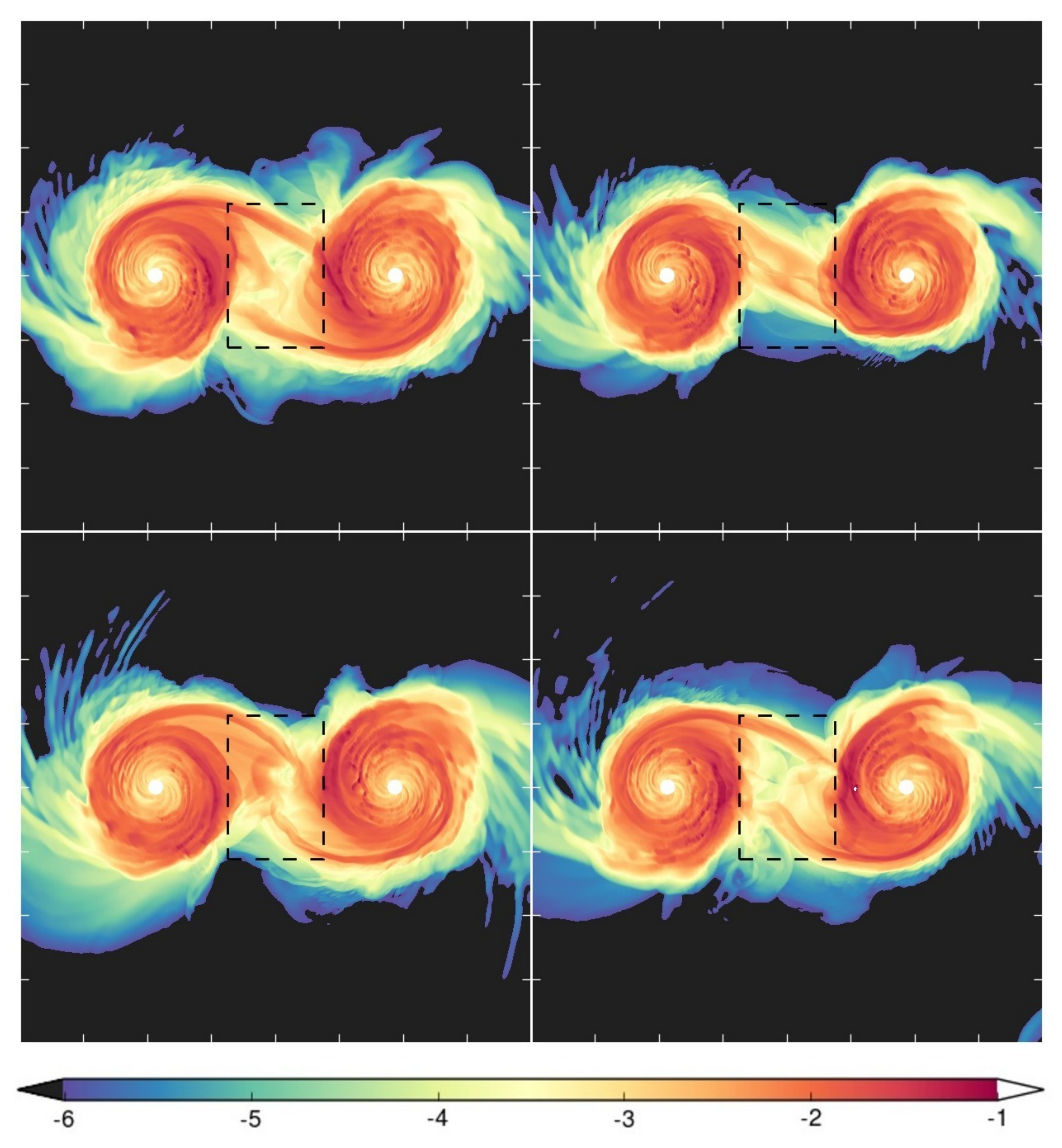}}
  \caption{Logscale density contours of the $20M$ run to demonstrate
    sloshing at various times; $5.55t_{\rm bin}$ (top left), $5.66t_{\rm bin}$ (top right),
    $5.91t_{\rm bin}$ (bottom left), and $6.03t_{\rm bin}$ (bottom right). We denote the
    ``sloshing region'' with the dashed black rectangle. At $5.55t_{\rm bin}$ we
    see two distinct arms connecting the BHs which then collapses to a
    single arm at $5.66t_{\rm bin}$. We then see the formation of a double armed
    stream pattern again at $5.91t_{\rm bin}$ and $6.03t_{\rm bin}$ corresponding to time
    lapses of $0.36t_{\rm bin}$ and $0.48t_{\rm bin}$. }
  \label{fig:sloshing-contour}
\end{figure*}

As previously remarked in Section~\ref{sec:results_overview}, matter
sloshes back and forth within a bar-like region centered on the binary
center-of-mass. Its mass during early times of our simulations
is largely due to a transient whose origin lies in our only
approximately hydrostatic initial condition for the mini-disks.
That this should be so is demonstrated most dramatically by the
$50M$ and $100M$ binary separation runs, in which the ratio of
sloshing mass to total mass quickly drops from its initial peak as
gas settles back onto the mini-disks (see
Figure~\ref{fig:sloshing-mass}).  However, if we consider the
mass in this region only after decay of the transient, we find a
remarkable increase as the separation diminishes below
$\simeq 50M$.  To quantify this statement, we must first define this
region more precisely: for our purposes, it is a rectangle in the
corotating frame with dimensions $\left(\tilde{x} \times \tilde{y}\right) =
\left(0.4a(t) \times 0.6a(t)\right)$, centered on the center-of-mass.
Because the separation $a$ in the initially $20M$ separation run
decreases with time, we adjust this box size as a function of time so
that its dimensions are always the same fraction of $a(t)$.  The
motivation for choosing these dimensions can be seen by inspection of
Figure~\ref{fig:sloshing-contour}.

Employing this definition, we can calculate the ratio $M_{\rm slosh}/M_{\rm cav}$ 
as a function of time measured in binary orbital
periods for the $20M$, $30M$, $40M$, $50M$ Large, and $100M$ Large hydrodynamic
runs. Here $M_{\rm slosh}$ is the mass within the box just defined,
while $M_{\rm cav}$ is the mass within a circle of radius $1.5a(t)$
from the center-of-mass.  This quantity is plotted in Figure~\ref{fig:sloshing-mass}.
All the different separation runs begin with similar values of this
ratio, $\simeq 5$--$10 \times 10^{-3}$. In the quasi-Newtonian $100M$
and $50M$ separation runs, this ratio drops to $\simeq2$--$8\times10^{-4}$ 
within three binary orbits. In sharp contrast to
this behavior, as the simulations become progressively more
relativistic, the mean fractional gas content of the sloshing region
increases monotonically. In the $20M$ separation run, this ratio
maintains its initial value--$\simeq 5 \times 10^{-3}$--albeit with factor of several
fluctuations, for the entire 14~orbit duration of the simulation.  In
other words, a decrease in separation by about a factor of 2 leads to
an increase in $M_{\rm slosh}/M_{\rm cav}$ of an order of magnitude.
The most natural explanation for this dramatic change is also the
progressively shallower gravitational potential found as the system
becomes increasingly relativistic (see Figure
\ref{fig:potential_lines}).
As a result, gas can more easily flow out of the potential well of a
single BH into the center-of-mass region.  However, it cannot remain
there in a stationary state because the L1 region is dynamically
unstable. Once in the sloshing region, gas can only move back and
forth between the relatively stable regions near the mini-disks.  Upon
reaching the truncation radius of a disk, the streams shock.

To gain further insight into the properties of the sloshing region, 
in Figure~\ref{fig:sloshing-contour} we plot density contours
at four selected times. This figure demonstrates that the
back-and-forth motions are concentrated into discrete streams, but
sometimes there are two streams at once, while at other times there is
only one. Due to the $q=1$ symmetry of our system, the internal motions of
the dual stream features are equal in magnitude and opposite one another.
We speculate that the number of streams is related to interaction with
the spiral structure (see Sec.~\ref{sec:spiralshocks}), which exhibits
both $m=1$ and $m=2$ components.

The stream pairs carrying mass between the mini-disks appear to form in a
quasi-periodic manner. To characterize this behavior quantitatively, we first pre-whiten
the $a=20M$ data by removing any secular linear trends in $\tilde{M}_{\rm slosh}(t)
\equiv M_{\rm slosh}(t)/M_{\rm cav}(t)$.  We calculate this
pre-whitened function ($\psi$) as
\begin{equation}
  \psi(t) = \frac{ \tilde{M}_{\rm slosh}(t) - \tilde{M}_{\rm fit}(t) }{ |M_{0}|}
\end{equation}
where $M_{0}$ is the largest value of $\tilde{M}_{\rm slosh}$,
$\tilde{M}_{\rm fit}(t)$ is a linear fit for $\tilde{M}_{\rm slosh}(t)$, 
and we use only data for $2000 \leq t/M \leq 6000$. This
corresponds roughly to 3.5--11 binary orbital periods.

\begin{figure}[htb]  
  \centerline{\includegraphics[width=\columnwidth]{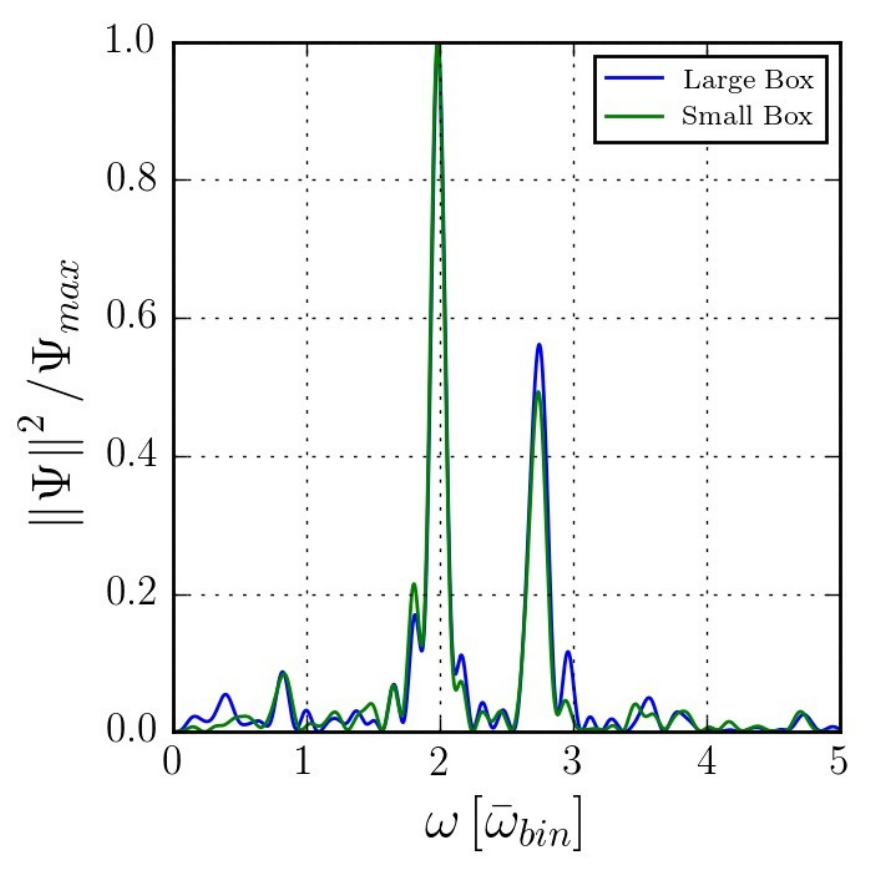}}
  \caption{Fourier power density of
    $M_{\rm slosh}(t)/M_{\rm cav}(t)$ for two regions in the
    $20M$ run. The ``Large Box" is centered on the center-of-mass and has dimensions $0.4a \times 0.6a$;
    it extends from the near-side of the BH1 mini-disk to the near-side of the BH2 mini-disk.
    The ``Small Box" is likewise centered on the center-of-mass, but has dimensions $0.26a \times 0.6a$.
    Angular frequency is measured in units of the mean binary orbital angular frequency during the time
    period covered by the Fourier analysis.}
  \label{fig:sloshing-fft}
\end{figure}

We then compute $\Psi$, the Fourier transform of $\psi$. In Figure~\ref{fig:sloshing-fft}
we plot $|\Psi|^2$, the Fourier power density, as a function of frequency.  It possesses
two distinct peaks, with angular frequencies roughly 2 and 2.75 times the mean orbital frequency of the
binary $\left(\bar{\omega}_{\rm bin}\right)$.  The width of these peaks is comparable to the range
of binary orbital frequencies during the period included in the Fourier analysis, from $\simeq 1.05\times$
the initial frequency to $\simeq 1.23 \times$ that frequency.

\subsection{Spiral Density Waves}
\label{sec:spiralshocks}
\begin{figure*}[htb]
  \centerline{\includegraphics[width=\textwidth]{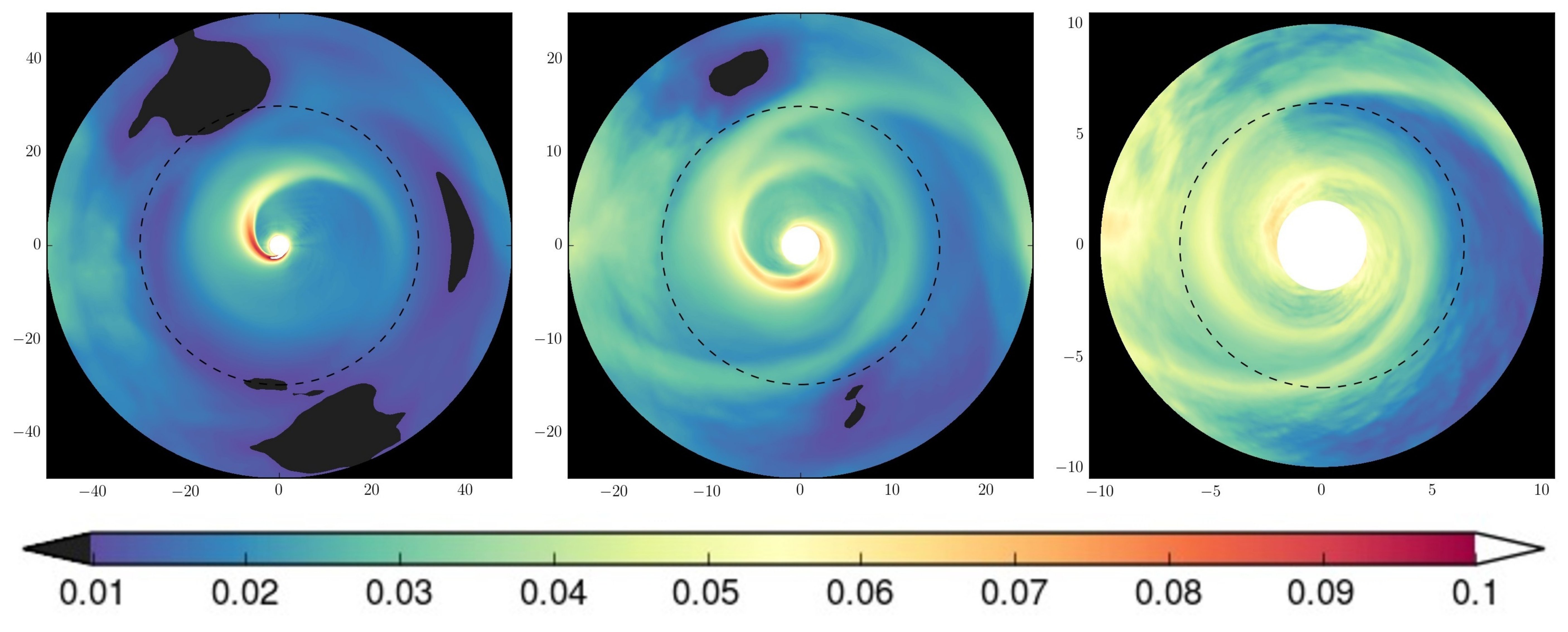}}
  \caption{(Left to right) Time-averaged local sound speed at binary separations of $100M$, $50M$, and $19.5M$. The white circle denotes the BH horizons while the dashed black line approximately corresponds to the tidal truncation radii of $0.3a$, $0.3a$, and $0.33a$ respectively.}
  \label{fig:spiral_cs}
\end{figure*}

Within the main body of the mini-disks, we observe the formation of spiral density waves.
They can be seen in density plots (Figures~\ref{fig:rho_snapshots}, \ref{fig:averaged_rho},
and \ref{fig:sloshing-contour}) and stand out even more clearly in a plot showing the local sound speed 
(Figure~\ref{fig:spiral_cs}) because a strong shock raises the local temperature by a much larger factor 
than it increases the density. These spiral density waves are of special interest, 
in part because they have the potential to alter the locally-emitted spectrum, but 
also because they may contribute significantly to accretion stresses.
In fact, spiral density waves were put forward very early on as a candidate 
mechanism to explain {\it all} accretion stresses~\citep{Lynden-Bell-Pringle74}, although 
subsequent studies determined that they were unlikely to be a general answer to the problem 
(see~\citep{Shu81,Papaloizou95,Boffin01} and references therein). 
In regions where the spiral density wave pattern speed is less than the local fluid orbital frequency 
of the disk, the spiral density wave orbits {\it backwards} through the disk, 
thereby carrying negative angular momentum with respect to the fluid in the disk. 
Once the spiral density wave amplitude becomes nonlinear, it can steepen into shocks,
allowing it to couple to the disk fluid via 
dissipation~\citep{Papaloizou95,Goodman2001,Heinemann2012,Rafikov16}. The resulting 
loss in the matter's angular momentum introduces a radial component to the flow and transports angular 
momentum outwards in the disk, thereby facilitating accretion onto the BH.

In accretion disks in binary systems, these spiral density waves are the 
result of the perturbing non-axisymmetric gravitational potential of the binary companion 
(see e.g.~\cite{SPL94} and references therein). The density waves in the disk due to the 
perturbing gravitational potential are launched at Lindblad 
resonances~\citep{Goldreich1979,PapaloizouLin1984}.
In Newtonian gravity, the perturbing potential (in a coordinate system centered on the 
central mass surrounded by the disk) due to the binary companion on a circular orbit 
is given by:
\begin{equation}\label{eq:indirect_potential}
\Phi_{\rm pert} = -\frac{M_2}{|\boldsymbol{r}-\boldsymbol{r}_2|} 
+ \frac{M_2 (\boldsymbol{r}\cdot\boldsymbol{r}_2)}{r^3_2},
\end{equation}
where $M_2$ is the mass of the perturber, $\boldsymbol{r}$ is the position vector 
and $\boldsymbol{r}_2$ is the position vector of the secondary (see, e.g.~\citep{Binney1987,SPL94}).
The first term in Eq.~\eqref{eq:indirect_potential} is the Newtonian potential of the 
secondary; the second term accounts for centrifugal forces due to the fact that
the corotating frame is non-inertial.

To analyze the effects of the perturbing potential, it is useful to expand it as a 
Fourier series in the azimuthal angle.  The $m$-th component of the perturbing potential excites
a spiral density wave with $m$ arms. If the perturber's orbit is not too close to the 
outer edge of the disk, the dominant azimuthal Fourier component of the Newtonian perturbing 
potential is the $m=2$ mode~\citep{SPL94}.  Consequently, in Newtonian simulations of these 
systems, a two-armed spiral density wave is often found to develop in the disk 
(see e.g. the simulations of~\citep{SPL94,Makita00,Ju16} or the early evolution of the 
disks studied in~\cite{KPO08}).

Because the tidal perturbation remains constant in the frame corotating with the binary,
the induced spiral density waves' pattern speed is the binary orbital frequency, 
which is slower than the orbital velocity of the disk fluid. 
In the fluid frame, a mode of order $m$ creates a spiral wave whose phase speed
is $m$ times the binary orbital frequency, as the fluid encounters 
$m$ spiral arms in each orbit.

Although Newtonian simulations find the development of clearly visible 
two-armed spiral density waves as the result of the dominant $m=2$
mode in the perturbing potential, our simulations show a more 
complicated flow morphology in the mini-disk for all binary 
separations studied.  To study the structure of the spiral waves in our
relativistic simulations, we compute the amplitude of different $m$ modes
in the disk surface density as a function of time (see, e.g.~\citep{Zurek1986,Heemskerk1992}):
\begin{equation}
D_m(t) = \int_{r_{\rm min}}^{r_{\rm max}} \, dr_{BL}d\phi_{BL} \Sigma(r_{BL},\phi_{BL};t)\, e^{-i m \phi_{BL}} \sqrt{-g_{BL}},
\end{equation}
where $r_{\rm min} = 0.13a$ and $r_{\rm max} = 0.38a$. The results are insensitive to 
the selection of $r_{\rm min}$. As a check, we performed this azimuthal mode analysis in two frames, 
a frame co-moving with BH1, and a frame rotating at the binary orbital period. Both calculations
give the same mode strengths, as expected.

\begin{figure}[htb]
  \centerline{\includegraphics[width=\columnwidth]{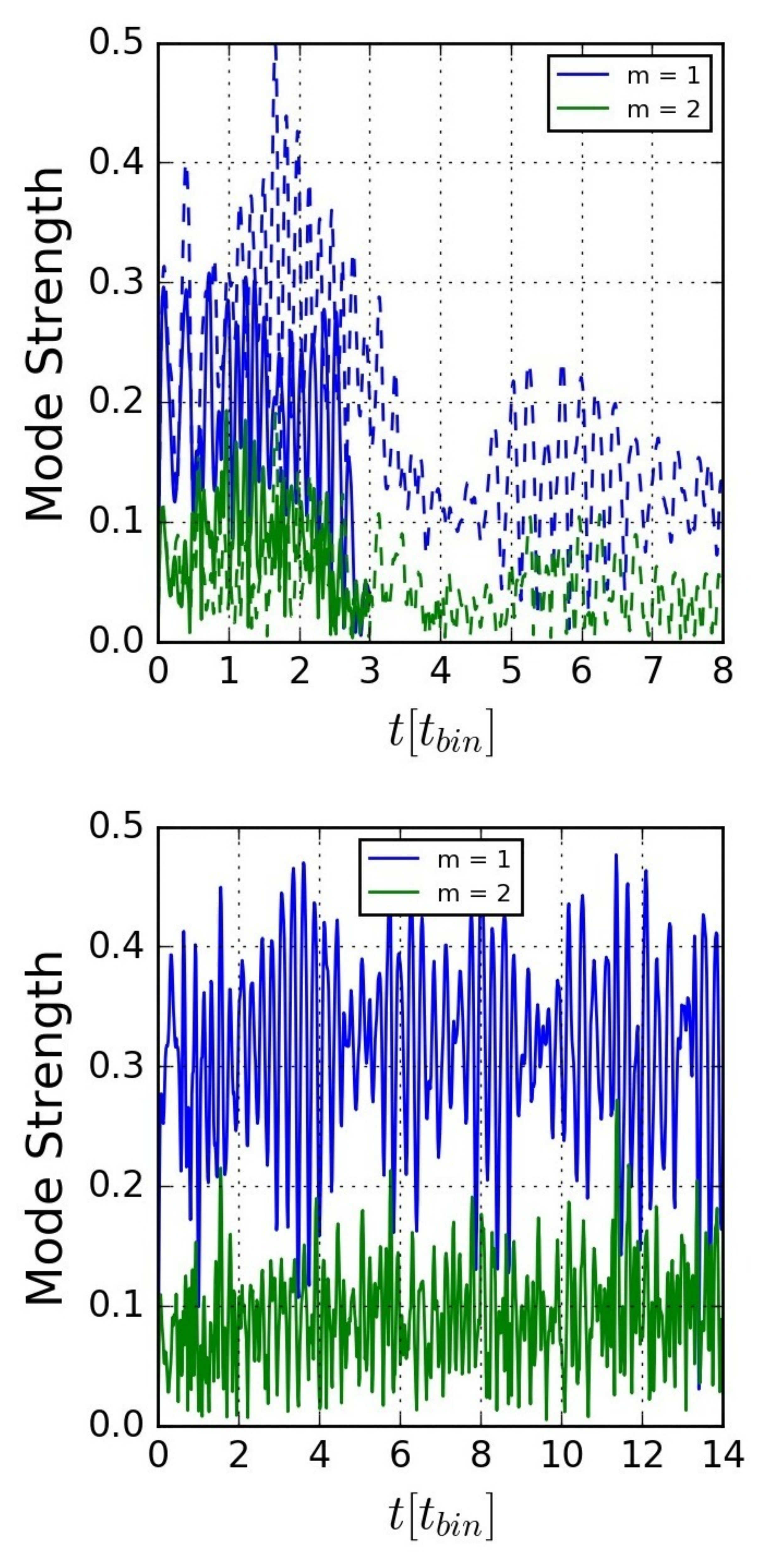}}
  \caption{Time-dependence of the $m=1$ and $m=2$ modes of the surface density 
  in the mini-disks.  (Top) Initial binary separation $100M$ shown by solid curves, $50M$ by dashed curves.
  (Bottom) Initial binary separation $20M$.}
  \label{fig:mode_analysis}
\end{figure}

The time-dependence of these mode amplitudes for initial binary separation $20M$, $50M$, and
$100M$ are shown in Figure~\ref{fig:mode_analysis}.  The $m=1$ amplitude is larger than the
$m=2$ amplitude in all cases, although $D_1/D_2$ appears to diminish over time for
the larger separations.  At late times in the $a=50M$ run, $D_1/D_2 \simeq 3$.
When the separation is as small as $20M$, $D_1/D_2$ oscillates around a constant mean value $\simeq 4$.
Although the $100M$ case did not run long enough to reach ``late times", $D_1/D_2$ appears to be
roughly $\sim 2$, suggesting a slow increase in this ratio with decreasing separation.

The origin of this departure from Newtonian experience appears to lie in the different
structure of the tidal potential in the PN regime.   To demonstrate this,
we computed the amplitudes with respect to $m$ in the potential in much the same fashion
as we did for the surface density; we call them $S_m$. In order to calculate them,
we used the same PN approximation for the perturbing potential as we did when
discussing the tidal potential itself, i.e., $\Phi = -(1 + g_{tt})/2$.  However, before
computing the mode integrals, we subtracted off the isotropic Newtonian potential due to
the individual BH.  The Newtonian ratio $S_1/S_2 = 0.143$,
irrespective of binary separation, because Newtonian gravity is scale-free.  However, relativistic
gravity is scale-dependent (see Table~\ref{tab:potential_modes}).  Even at $a=100M$,
$S_1/S_2$ is close to triple the Newtonian value, and it grows monotonically with decreasing
binary separation, exceeding unity for separations slightly less than $30M$.  Relativistic
contributions to $m=1$ modes can be significant at separations as large as $100M$ because
they first appear in the PN expansion in the lowest-order non-Newtonian terms
(see, e.g., the explicit expression for the PN Lagrangian governing a test-particle
in a binary system in \cite{2PNROCHE}).  In addition, the same low-order terms also create
new contributions to $m=2$ modulation, altering the Newtonian component in both magnitude
and phase.

\begin{deluxetable}{l c}
\tablecolumns{2}
\tablewidth{0.75\columnwidth}
\tablecaption{Azimuthal Potential Modes \label{tab:potential_modes}}
\tablehead{
\colhead{$a[M]$}      & \colhead{$S_{1}/S_{2}$}
}
\startdata
Newtonian       & 0.143   \\ 
100             & 0.412   \\ 
50              & 0.665   \\ 
40              & 0.825   \\ 
30              & 0.990   \\ 
19.5            & 1.209   \\ 
19.0            & 1.212   \\ 
18.5            & 1.239   \\ 
18.0            & 1.276   \\ 
17.5            & 1.298   \\ 
17.0            & 1.335   \\ 
16.5            & 1.359   \\ 
16.0            & 1.382   
\enddata

\tablecomments{Strength of the $m=1$/$m=2$ modes of the perturbing potential as a function of separation.}

\end{deluxetable}

As a final point, we note that we have performed a test simulation of a BBH system
in which initially only BH1 was surrounded by a mini-disk, in order to test
if shocks induced by the sloshing might produce spiral density waves in the mini-disks. 
The morphology of the fluid flow and mode strengths were very similar to those from 
simulations with two mini-disks. We therefore conclude that the 
main driving factor of the spiral waves in our mini-disks is indeed the tidal field of the
binary companion.

\section{Discussion}
\label{sec:discussion}

\subsection{Tidally Truncated Mini-Disks}
\label{sec:discussion_tidaltruncation}

Previous discussions of the shapes of disks in binary systems have been almost exclusively posed in terms
of Newtonian gravity.  In addition, when describing their size, it has generally been customary to specify
only a mean radius from the mass at the center of such a disk, often without a precise operational definition
of what exactly makes that radius the \enquote{size} of the disk.  In the course of our investigation of PN effects
on the tidal truncation of disks in binaries, we have found that quantitative description demands such a
definition: these disks are, at the $\sim 20\%$ level, non-circular even in the Newtonian limit, and
their sizes (when more precisely defined) do begin to change as gravity departs from Newtonian.

In principle, there could be many definitions of the \enquote{size} of a disk in a binary system.  Here we explored
the implications of two choices.  One, a definition focusing on the surface mass density profile, is suited
to answering questions such as ``if an external fluid stream strikes the disk, where does it encounter the
full inertia of the disk?"  The other, a definition focusing on the shapes of disk material's orbits, is
designed to discriminate between fluid elements that repeatedly orbit the central mass of a disk from those
that may pass near the disk, but then swing away from it.

Even in the Newtonian limit (exemplified by our $100M$ separation simulation), the disks appear to be
asymmetric in terms of both these measures.  The far side (away from the other mass in the binary system)
extent defined by the surface density gradient agrees with the near side extent defined by streamlines,
and both are consistent with the usual Newtonian estimate for equal-mass binaries ($\simeq 0.30a$), but
the near side surface density definition and the far side streamline definition both give extents $\sim 20\%$
smaller.  In other words, by the surface density definition, disks are cut off more tightly on the near
side than the far, while by the stream line definition they extend farther from the central mass on
the near side.   Thus, at this level of precision, it's clear that a single \enquote{truncation radius} fails to be
an adequate description: the disks are neither round nor admit a single definition appropriate to all
intuitive meanings of \enquote{disk edge}.

Relativistic effects also begin to alter mini-disk shapes once $a$ is a few tens of $M$ or less, particularly
with respect to the near-side surface density definition.  On the near-side, the disk extends to a progressively
larger and larger fraction of $a$ as the binary separation moves farther and farther into the relativistic
regime.  When $a \simeq 20M$, the (surface density) near edge is $\simeq 35\%$ farther from the central mass than in the
Newtonian limit.  As we have already remarked, this effect can be attributed directly to the progressively
shallower potential gradient in the L1 region produced by increasingly relativistic gravity (as illustrated in
Figure~\ref{fig:potential_lines}).  When the binary mass-ratio is less than unity, the relativistic effects are,
if anything, stronger (Figure~\ref{fig:potential_lines_qneq1_secondary}).  A similar result, also in the PN formulation
but posed in terms of the volume occupied by the secondary's Roche lobe, was found by \cite{2PNROCHE}.  It is
worth noting, however, that for their parameters (separations of $5M$ and $10M$), the
Roche lobe volume decreases with greater relativistic effects when $q < 0.7$.

On the basis of our results, we might speculate that at still smaller separations the mini-disks continue to exist until the
truncation radius (defined in any of these senses) becomes only somewhat larger than the absolute minimum scale
set by the ISCO.  However, once the inspiral becomes more rapid than the internal accretion rate within such
a disk, matter from the outer edge, which now extends beyond the truncation radius, must lose its binding to
an individual BH, and instead travel through both its original Roche lobe and the partner's.  Such
effects are closely related to the topic of the next subsection, the ``sloshing region".

\subsection{Sloshing}
\label{sec:discussion_sloshing}
 
Gas extending across the L1 region in a system with binary disks has
been previously observed in stellar systems \citep{Mayama10}; it has
also appeared in Newtonian hydrodynamics simulations of binary stellar
systems \citep{VB11,Nelson16} and SMBBHs \citep{Farris14,DOrazio16},
although without comment in the latter work.  However, we have found a
number of new properties of this gas.  It is in a constant state of
back-and-forth motion, and, as shown in Figure~\ref{fig:sloshing-fft},
the variations in this motion are strongly modulated at two
frequencies, one very close to $2\bar{\omega}_{\rm bin}$, the other at
$2.75\bar{\omega}_{\rm bin}$.  Second, and most surprisingly, the
fraction of disk mass in this sloshing region rises steeply as
relativistic effects become important.

The quantity of gas in the sloshing region is determined through a
complex mechanism. When sloshing gas shocks against the edge of a
mini-disk, it ``spalls" off gas parcels, giving them enough energy to
travel across the L1 region.  The amount required is comparatively
small.  Although Figures~\ref{fig:potential_lines},
\ref{fig:potential_lines_qneq1_secondary}, and \ref{fig:potential_lines_qneq1_primary}
show the effective gravitational
potential in the rotating frame of the binary, they do not include the
contribution to the effective potential associated with the orbital
motion of a fluid element around one of the BHs. Even in the
Newtonian limit, this contribution causes the total effective
potential to be rather shallower than the curves shown in this figure;
when PN effects further decrease the change in effective potential
from the edge of the mini-disk to the L1 point, the energy delivery
threshold for liberation of gas becomes only a small fraction of
$GM/a$. In agreement with this potential picture, the mean fraction of
available gas in the sloshing region ($M_{\rm slosh}/M_{\rm cavity}$)
increases monotonically with decreasing binary separation, with a
particularly sharp increase for $a \lesssim 30M$. In addition,
animations of our simulations show correlations between fluctuations
in spiral wave structure and ejection of matter into the sloshing
region.  It is possible that when accretion streams from a
circumbinary disk enter the picture, their impact may also influence
gas injection into the sloshing region.

These new properties may lead to a significant new observable.  Every time a sloshing stream shocks against the edge of a
mini-disk, the associated energy dissipation is available for photon radiation.  Moreover, the periodic character of these motions
means that if the cooling time post-shock is shorter than the dynamical period, the radiation may be strongly modulated.  Because the
characteristic speed of this motion is comparable to the orbital speed, and therefore increases $\propto a^{-1/2}$, and the amount
of mass involved increases much more rapidly with decreasing binary separation when $a \lesssim 30M$, the energy available for
this potentially periodic radiation signal should increase sharply as the binary inspirals through the last few tens of $M$ in
separation.

More quantitatively, for BBH systems with mass-ratios of order unity, we may estimate the time-averaged
heat release in the sloshing region by
\begin{equation}
L_{\rm slosh} \sim \left(M_{\rm slosh}/M_{\rm cav}\right) \left(\dot M t_{\rm in}\right) v_{\rm orb}^2 \omega_{\rm slosh},
\end{equation}
where the total mass of the mini-disks $M_{\rm cav}$ is determined by the accretion rate $\dot M$ and the inflow time
through an individual mini-disk $t_{\rm in}$.  We further estimate the speed of the sloshing to be comparable to the
binary orbital speed $v_{\rm orb}$.  The repetition rate of the sloshing cycle is $\omega_{\rm slosh}$, which we have already
determined to be $\simeq 2 \omega_{\rm bin}$.  For greater insight, this luminosity estimate may be rewritten as
\begin{equation}
L_{\rm slosh} \sim \left(M_{\rm slosh}/M_{\rm cav}\right) (r_g/a) \left[(h/r)^2 \alpha^\prime\right]^{-1}  L_{\rm acc},
\end{equation}
in which $L_{\rm acc}$ is the luminosity released by accretion onto the BHs, $h/r$ is the aspect ratio of the
mini-disks, and $\alpha^\prime$ is the ratio between vertically-integrated
and time-averaged accretion stresses and the similarly vertically-integrated and time-averaged disk pressure.  We
write it as $\alpha^\prime$ rather than the conventional $\alpha$ as a reminder that stresses associated with spiral
waves can add to the usual correlated MHD turbulence.  We have already found that when $a=20M$, $M_{\rm slosh}/M_{\rm cav} \simeq
3 \times 10^{-3}$.  Although the disk aspect ratio must certainly depend on such parameters as the accretion rate,
for the time being we simply note that if the accretion rate is not far below Eddington, when the system is in the
PN regime, $h/r \sim 10^{-2}$ and $\alpha^\prime \sim 10^{-1}$ should be reasonably conservative estimates.
If so, $L_{\rm slosh}$ might be somewhere in the neighborhood of $\sim 10^{-1}L_{\rm acc}$.  Further work will be
required to make sensible predictions about its spectrum.

The optical depth in the sloshing region can be estimated in similar fashion:
\begin{equation}
\tau_{\rm slosh} \sim 4 \left(M_{\rm slosh}/M_{\rm cav}\right) (r_g/a)^{1/2} \left[(h/r)^2 \alpha^\prime\right]^{-1} (\dot m/\eta),
\end{equation}
where $\dot m$ is the ratio of the accretion rate to the Eddington rate and $\eta$ is the rest-mass efficiency of energy
release by accretion.  Here, to be consistent with the definition of $M_{\rm slosh}$ found in Sec.~\ref{sec:sloshing}, we estimate
the area occupied by the sloshing region as $\simeq a^2/4$. Repeating the estimate we just made for the mini-disk internal inflow rate,
this expression suggests that the sloshing region should generally be optically thin to Thomson scattering because
$(r_g/a)^{1/2} \lesssim 1$ while $\dot m/\eta$ is likely to be bounded above by $\sim 10$, but could be considerably less.
Thus, in many instances the $\sim 10\%$ addition to the time-averaged bolometric luminosity should, in fact, be modulated
with the period of the sloshing, half the  binary orbital period.

The sloshing may also lead to a wholly new kind of mass-transfer.  In our simulation, with its unity mass-ratio, the sloshing is,
on average, perfectly symmetric.  However, the general case for binaries is a mass-ratio different from unity; when this is true,
the sloshing is likely to lose its symmetry, so that there is a net mass-flow from one mini-disk to the other.  As a result, mass
that leaves the inner edge of the circumbinary disk and arrives at the outer edge of the mini-disk around, for example, the secondary,
may, through the sloshing mechanism, find its way to the mini-disk around the primary. This will be an interesting phenomenon for
future simulations to explore.

\subsection{Spiral Density Waves}
\label{sec:discussion_spiralshocks}

Perhaps somewhat surprisingly, although spiral waves in disks within binaries were first studied
in the context of cataclysmic variables \citep{Lynden-Bell-Pringle74}, disks in SMBBHs
whose separations are small enough to be in the relativistic regime may be the environment in
which their effects are strongest.   The reason is that
the angular momentum transport that can be accomplished by spiral shocks increases with
disk temperature, in part because a higher ratio of disk sound speed to orbital speed
makes the spiral opening angle larger, leading to stronger shocks \citep{SPL94,Ju16};
disks in SMBBHs are especially hot in this sense.
In such disks, if the accretion rate in Eddington units $\dot m\gtrsim 0.01$, the
local pressure is dominated by radiation out to $\sim 100M$, and the ratio of the
effective sound speed (including radiation pressure) to orbital speed is $ \gtrsim 0.1$ 
for all radii within $\sim 15 (\dot m/0.1) M$ \citep{SS73}. Thus, it is possible that spiral shocks, driven by some 
combination of tidal forces and accretion stream shocks, may be of 
special interest in relativistic SMBBHs.  We caution, however, that there are 3-dimensional effects which 
might complicate the situation (see Sec.~\ref{sec:discussion_3d} for further discussion of this point).
We also remark that our simulations described disk thermodynamics in an extremely simplistic fashion, so
the details of spiral shock behavior in them should not be taken as predictive of real systems.

The spiral shocks in relativistic SMBBHs may be enhanced in another way as well.
As the binary peels streams of gas off the inner edge of the circumbinary disk, 
a portion of the streams falls into the central cavity and shocks against the outer 
edges of the mini-disks. These shocks could also contribute to spiral shocks in two ways. 
First, the accretion stream shocks may heat the mini-disk and therefore enhance the 
effectiveness of angular momentum transport in spiral shocks. The degree to which the 
temperature throughout the mini-disk is raised by the accretion shocks will depend on 
how rapidly the heat they generate is radiated.  Some estimates~\citep{Roedig:2014} 
suggest that this reradiation is quite rapid, but the immediate post-shock temperature 
is so high that the ultimate post-shock temperature might still exceed the temperature 
expected from ordinary disk dissipation processes. Secondly, other 
simulations~\citep{Shi12,Noble12,DOrazio13,Farris14} have shown that the rate at which 
matter is delivered from a circumbinary disk to mini-disks exhibits strong periodic 
modulation, with a period comparable to the binary orbital period. This periodicity, 
which does not affect ordinary mass transfer through Roche lobe overflow, could drive 
the formation of additional spiral shocks.

It is also of interest that spiral shocks in relativistic SMBBHs have a different symmetry
from those in Newtonian systems.  Previous Newtonian simulations (see, e.g.~\citep{SPL94,Makita00,Ju16}),
as well as analytic theory~\citep{SMIS87,SPL94,Rafikov16}, predict the dominant spiral shock mode 
should be $m=2$.  So, too, does the recent work of~\cite{RyanMacFadyen16}, in which mini-disk
dynamics driven by the BH at the center of the disk were treated in full general relativity,
but the tidal field due to a BH companion was added as a perturbation.  By contrast,
our work found that the amplitude of surface density modulation associated with $m=1$ modes
was always a few times greater than that due to $m=2$ modes.  This discrepancy is almost certainly explained
by the fact that in all these previous efforts the companion's tidal field was described in terms of Newtonian
gravity, as in Eqn.~\ref{eq:indirect_potential}.  In this context, it should be noted that the
Newtonian approximation was actually appropriate to the work of \cite{RyanMacFadyen16} because the
binary separation in their simulations was fixed at $1000M$.

In fact, not only do the $m=1$ modes dominate $m=2$ for separations as large as $100M$, the ratio of
$m=1$ to $m=2$ grows as the binary separation decreases and general relativistic effects become
even more prominent.  The ratio might increase still further at very small separation
because inspiral, whose timescale is proportional to $a^4$, is another source of $m=1$ behavior.
The inspiral rate due to the loss of energy to GW emission depends
on the mass ratio of the binary, and is highest for equal mass binaries.
Therefore, the two BHs in the binary approach one another               
faster than their respective mini-disks; on their own, the mini-disks would inspiral 
much more slowly due to less efficient GW emission. As a result, at any given time the
BHs are slightly offset from the centers of their own disks, with the magnitude of
this offset comparable to the orbital shrinkage during a mini-disk fluid dynamical time.  Relative
to the size of the mini-disk, this offset can become sizable in the late stages of inspiral:
\begin{equation}
\frac{\Delta x}{r_t} \sim 0.1 \frac{q^{1/2}}{(1+q)^{3/2}} (r_t/0.3a)^{1/2} (a/10M)^{-5/2}.
\end{equation}
Analogs to this effect have been seen in other kinds of systems.  For example, an inverse
version can be observed in self-gravitating disks around single masses: in this case, the central mass
is forced onto a spiral trajectory by the gravitational potential of an $m=1$ mode in the disk
\citep{ARS89,Heemskerk1992,Korobkin11,Mewes16a,Mewes16b}.
Similarly, $m=1$ spiral shocks can be created in accretion disks around black 
holes recoiling after a merger~\citep{Corrales2010,Ponce2012}.

\begin{figure}[htb]
  \centerline{\includegraphics[width=\columnwidth]{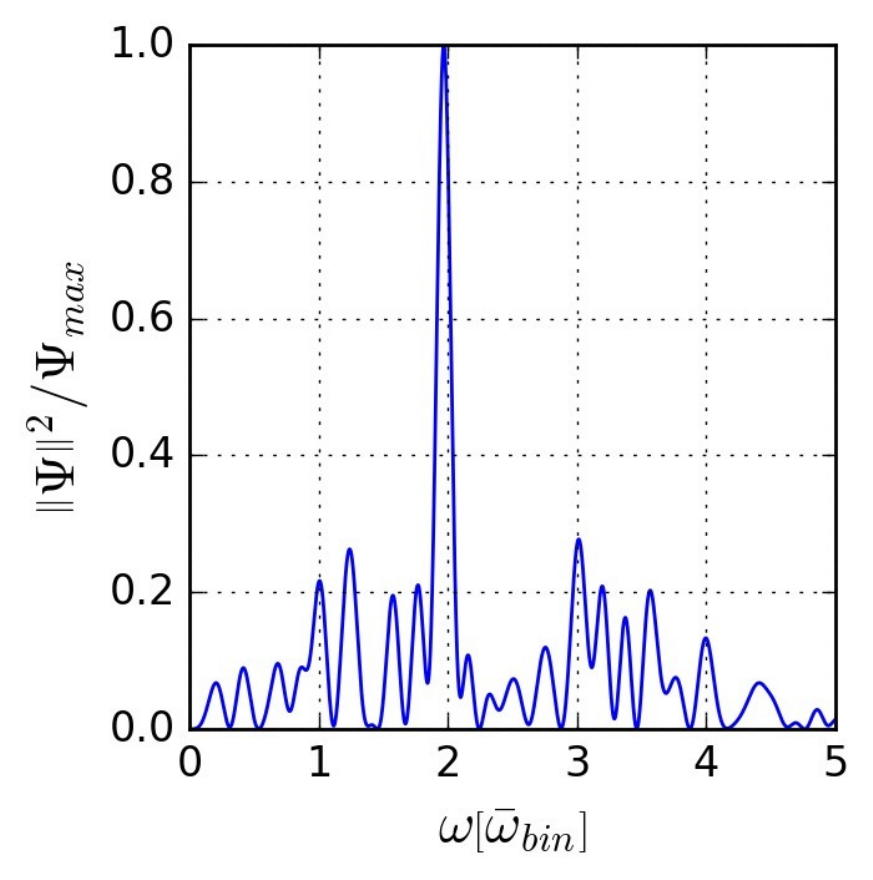}}
  \caption{Fourier power density of the accretion rate onto BH1 as a function of angular frequency measured
   in units of the binary angular frequency.}
  \label{fig:mdot-fft}
\end{figure}

We close this section by pointing out a curious link between spiral shocks,
accretion onto the BHs, and sloshing.  As shown in Figure~\ref{fig:sloshing-fft},
the mass in the sloshing region is modulated at two frequencies, $2\omega_{\rm bin}$
and $2.75\omega_{\rm bin}$.  Surprisingly, we also find (see Figure~\ref{fig:mdot-fft}) that the accretion rate
onto BH1 (and presumably its twin, BH2) is {\it also} modulated, but exclusively at frequency
$2\omega_{\rm bin}$.  We emphasize that this is {\it not} the modulation associated with accretion
from the inner edge of the circumbinary disk \citep{Shi12,Noble12,DOrazio13,Farris14}; it is
the accretion rate inside a mini-disk wholly isolated from any external circumbinary disk.

In our purely hydrodynamic simulations, the only way matter can accrete through a mini-disk
is by stresses associated with spiral shocks.  However, to the extent that these spiral
shocks are well described by their time-averaged structure, there is no reason for them
to cause any time-dependence in the accretion rate because, on average, they are stationary
in the corotating frame.  It is possible, though, that their strength, and therefore the
accretion rate, might be modulated due to interaction between the dominant $m=1$ and the
weaker $m=2$ mode.   We further speculate, prompted by the $2\omega_{\rm bin}$ modulation
of the sloshing mass, that this periodic variation of the spiral shock amplitude is related
to periodic behavior in the sloshing region.  Some of the angular momentum carried outward by the
spiral shocks may be given to sloshing mass; conversely, periodic motions in the sloshing
region may drive corresponding changes in amplitude in the spiral shocks.

\subsection{Neglected Three-Dimensional Effects}
\label{sec:discussion_3d}

All our calculations were confined to 2D dynamics in the disk equatorial plane.  Real disks
are, of course, 3D objects.  Although many effects we investigated (e.g., the location of
the disks' outer edges) are unlikely to be substantially affected by a change in dimensionality,
others are.

Two in particular are worth comment.  The first is that the thickness of the accretion
streams from the circumbinary disk to the mini-disks relative to the thickness of the
mini-disks themselves could be an important parameter regulating how the material in
these streams joins the disk and the character of waves launched into the disk by stream
impact.  For example, if the streams are thicker than the disks, how far inward do the
shocks wrap around the disks?  When the streams shock against a disk edge, do they rise
in temperature sufficiently that they are always thicker than the disk?  Similar questions
might also apply to the sloshing streams.

The second is that there can be significant 3D effects in the propagation of the spiral
waves within the mini-disks.  \citet{LubowOg98} and \citet{LubowOg99} studied this issue for the case of
waves driven by tidal gravity in a binary.  When the disk temperature decreases
vertically away from the midplane, they found that the spiral waves are channeled
toward the surface, but also that a more gradual drop in gas density at the surface can
limit the concentration of the waves there.  The degree of wave focusing can, of course,
have significant implications for where the waves steepen into shocks and dissipate their
energy.

Proper resolution of these questions must, of course, await fully 3D calculations.

\section{Conclusions}
\label{sec:conclusion}

In this paper, we have presented the first exploration of how 
mini-disks in binary systems behave when the binary separation is small
enough to make general relativistic effects in the spacetime,
particularly regarding tidal gravity, significant.

The gravitational potential along the line between the two masses
becomes shallower, and its gradient gentler, as the system becomes
more relativistic.  Within the secondary's Roche lobe, this
contrast between relativistic disks and Newtonian grows with mass-ratios
further from unity.  One result, apparent in our $q=1$ simulations, is that,
particularly as the separation becomes $\lesssim 30M$, the disks stretch toward the L1
point.  The resulting asymmetry is large enough that the outer rims of
disks in a relativistic binary are significantly non-circular, so that
thinking in terms of a ``tidal truncation radius" for such disks can
be misleading.

Newtonian studies \citep{Mayama10,Farris14,DOrazio16} had previously
shown that a small fraction of the disks' mass can be removed from the
individual disks within a binary and placed in the region stretching
from one disk to the other through the L1 point.  A further
consequence of the shallower gradient in the relativistic regime is a
sharp increase in that fraction, an order of magnitude increase
between binary separations of $50M$ and $20M$.  At this level, the
``sloshing mass" can play a significant role in the system.  For
example, when the mass-ratio is not unity, asymmetry in the sloshing
may create an entirely new way for mass to pass from one part of the
binary system to another.

This sloshing may also result in a striking and unique electromagnetic
signal of a BBH system in the period shortly before merger.  In
the regime of separations in which the sloshing mass is sizable, the
repeated shocks it suffers may account for $\sim 10\%$ of the
bolometric luminosity; in addition, in many circumstances the region
in which the heat is released may be optically thin enough for its
lightcurve to follow the periodic character of the heat release.  For
the $q=1$ case, there are two frequency components in the modulation,
one at twice the binary orbital frequency and another at $\simeq 2.75
\times$ that frequency.  For binary separation $20M$, these correspond
to periods $\simeq (1/2) M_6$~hr, for $M_6$ the total binary mass in
units of $10^6 M_\odot$.

We have also discovered that relativistic alteration of the tidal
forces leads to other contrasts with Newtonian behavior.  It has long
been known that tidal forces can drive spiral waves in disks within
binary systems; in the Newtonian limit, these have exclusively $m=2$
character.  On the other hand, even the lowest-order relativistic
corrections can introduce $m=1$ perturbations into the binary
potential while also altering the $m=2$ component.  Higher-order terms
such as those associated with gravitational radiation and the orbital
evolution it creates can also lead to new $m=1$ and $m=2$ components
in disk dynamics.  In consequence, the $m=1$ component can become the
dominant feature even when the separation is as large as $\sim 100M$.

Whether $m=1$ or $m=2$, spiral shocks can supplement the angular
momentum transport produced by MHD stresses.  Future work will
determine the degree to which this transport is altered by the change
in spiral structure in the relativistic regime.  This will be a topic
of particular interest because the effectiveness of angular momentum
transport by spiral shocks increases with the ratio between the disk
matter's sound speed and orbital speed; when the accretion rate in an
SMBBH is close enough to Eddington to make relativistic regions
radiation-dominated, these relativistic disks may be particularly
strongly affected by such shocks.

As inspiral progresses to yet smaller separations, both of these
relativistic effects are likely to become stronger.  The ``softening"
of the potential can provide a channel for mass-loss from the
individual disks; their dissolution is likely to be further
accelerated when the orbital evolution time becomes shorter than the
inflow time within the disks.  To determine the consequences of these
processes requires further work along these lines.



\section*{Acknowledgments}
We thank Brennan Ireland for a careful reading of this
manuscript and for helpful discussions. We would also like
to thank Hiroyuki Nakano and Yosef Zlochower for helpful discussions.
J.~K. would like to thank
Caroline Terquem and John Papaloizou for a very informative conversation.
We would like to thank the anonymous referee for the careful reading of this
manuscript and for the helpful comments and questions raised.
D.~B., M.~C., V.~M., and S.~C.~N. received support from NSF
grants AST-1516150, AST-1028087, PHY-1305730, and ACI-1550436.  
J.~K. was partially supported by NSF grants AST-1028111 and AST-11516299.

Computational resources were provided by XSEDE allocation 
TG-PHY060027N and by the BlueSky Cluster at Rochester Institute of Technology.  
The BlueSky cluster was supported by NSF grants AST-1028087 and PHY-1229173.
This research was also part of the Blue Waters sustained-petascale computing
NSF projects ACI-0832606, ACI-1238993, and OCI-1515969, OCI-0725070. Blue Waters is
a joint effort of the University of Illinois at Urbana-Champaign and
its National Center for Supercomputing Applications.



\bibliography{bhm_references}

\end{document}